%% file: ms_revised2.tex
\documentclass[apj]{emulateapj}

\usepackage{mathptmx}
\usepackage{color}
\usepackage{subfigure}

\journalinfo{Draft version \today}
\slugcomment{ApJ in press}
\shorttitle{HIGHLY RADIO-LOUD QUASARS AT $z>4$}
\shortauthors{WU ET AL.}

\def\simgt{\lower 2pt \hbox{$\, \buildrel {\scriptstyle >}\over {\scriptstyle \sim}\,$}}
\def\simlt{\lower 2pt \hbox{$\, \buildrel {\scriptstyle <}\over {\scriptstyle \sim}\,$}}
\def\chandra{{\it Chandra}}

\def\rosat{{\it ROSAT}}

\def\xmm{\hbox{\it XMM-Newton}}
\def\wise{{\it WISE}}
\def\swift{{\it Swift}}

\def\civ{C~{\sc iv}}
\def\mgii{Mg~{\sc ii}}
\def\lyanv{\hbox{Ly$\alpha$ + N~{\sc v}}}

\def\pl{\hbox{power-law}}

\def\aox{$\alpha_{\rm ox}$}
\def\daox{$\Delta\alpha_{\rm ox}$}
\def\daoxrq{$\Delta\alpha_{\rm ox, RQQ}$}
\def\daoxrl{$\Delta\alpha_{\rm ox, RLQ}$}

\def\zfour{\hbox{$z>4$}}

\def\xray{\hbox{X-ray}}
\def\garay{\hbox{$\gamma$-ray}}

\begin{document}


\title{An X-ray and Multiwavelength Survey of Highly Radio-Loud Quasars at \boldmath $z > 4$: Jet-Linked Emission in the Brightest Radio Beacons of the Early Universe}


\author{Jianfeng~Wu\altaffilmark{1,2,3}, 
W.~N.~Brandt\altaffilmark{1,2},
Brendan~P.~Miller\altaffilmark{4},
Gordon~P.~Garmire\altaffilmark{1},
Donald~P.~Schneider\altaffilmark{1,2},
Cristian~Vignali\altaffilmark{5,6}
}

\altaffiltext{1}
     {Department of Astronomy \& Astrophysics, The Pennsylvania State
     University, 525 Davey Lab, University Park, PA 16802, USA}
\altaffiltext{2}
    {Institute for Gravitation and the Cosmos, The Pennsylvania State 
     University, University Park, PA 16802, USA}
\altaffiltext{3}
    {Harvard-Smithsonian Center for Astrophysics, 60 Garden Street, Cambridge, MA 02138, USA}
\altaffiltext{4}
    {Department of Astronomy, University of Michigan, Ann Arbor, 
     MI 48109, USA}
\altaffiltext{5}
    {Dipartimento di Astronomia, Universit\`{a} degli Studi di
     Bologna, Via Ranzani 1, I--40127 Bologna, Italy} 
\altaffiltext{6}
    {INAF -- Osservatorio Astronomico di Bologna, Via Ranzani 1, I--40127 Bologna, Italy}
\email{jfwu@astro.psu.edu}


\begin{abstract}
We present a systematic study of the \xray\ and multiwavelength
properties of a sample of 17 highly radio-loud quasars (HRLQs) at
\zfour\ with sensitive \xray\ coverage from new \chandra\ and archival
\chandra, \xmm, and \swift\ observations. Eight of the new and
archival observations are reported in this work for the first
time. New \chandra\ observations of two moderately radio-loud and
highly optically luminous quasars at \hbox{$z\gtrsim4$} are also
reported. Our HRLQ sample represents the top $\sim5\%$ of radio-loud
quasars in terms of radio loudness. We found that our HRLQs have an
\xray\ emission enhancement over  
HRLQs at lower redshifts (by a typical factor of $\approx3$), and this
effect, after controlling for several factors which may introduce
biases, has been solidly estimated to be significant at the
3--4$\sigma$ level. HRLQs at $z=3$--4 are also found to have
a similar \xray\ emission enhancement over $z<3$ HRLQs, which
supports further the robustness of our results. We discuss
models for the \xray\ enhancement's origin including a 
fractional contribution from inverse Compton scattering of cosmic
microwave background photons. No strong correlations are found between the
relative \xray\ brightness and optical/UV emission-line rest-frame
equivalent widths (REWs) for radio-loud quasars. However, the line
REWs are positively correlated with radio loudness, which suggests
that relativistic jets make a negligible contribution to the
optical/UV continua of these HRLQs (contrary to the case where the emission
lines are diluted by the relativistically boosted continuum). Our
HRLQs are generally consistent with the known \hbox{anti-correlation} between radio
loudness and \xray\ power-law photon index. We also found that the two
moderately radio-loud quasars appear to have the hardest
\xray\ spectra among our objects, suggesting that intrinsic \xray\
absorption ($N_{\rm H}\sim10^{23}\ {\rm cm}^{-2}$) may be present. Our
\zfour\ HRLQs generally have higher \xray\ luminosities than those for
the composite broad-band spectral energy distributions (SEDs) of HRLQs
at lower redshift, which further illustrates and supports the \xray\ emission
enhancement of \zfour\ HRLQs. Some of our HRLQs also show an excess of
mid-infrared emission which may originate from the synchrotron
emission of the relativistic jets. None of our \zfour\ HRLQs is
detected by the {\it Fermi} LAT two-year survey, which provides
constraints on jet-emission models.  

\end{abstract}

\keywords{galaxies: active --- galaxies: nuclei --- galaxies: high-redshift ---
jets: galaxies --- \hbox{X-rays}: galaxies}


\section{Introduction}\label{intro}

X-ray studies of \zfour\ active galactic nuclei (AGNs) over the past $\approx12$~yr  
have greatly improved understanding of the growth of the first 
supermassive black holes (SMBHs) in the early Universe, thanks to unprecedentedly 
powerful \xray\ observatories as well as wide-field surveys in the optical/UV and radio 
bands. \chandra\ and \xmm, along with other \xray\ missions, have observed more than 100
quasars at \zfour, the majority of which were 
first identified by wide-field optical/UV surveys, e.g., the Sloan Digital 
Sky Survey (SDSS; York et~al. 2000), or radio surveys, e.g., the Faint Images 
of the Radio Sky at Twenty-Centimeters (FIRST; Becker et~al. 1995)
survey and the NRAO VLA Sky Survey (NVSS; Condon et~al. 1998). The
quasar population is divided into radio-quiet and radio-loud
subcategories based on the apparently bi-modal distribution of the
radio-loudness parameter (e.g., Ivezi\'{c}
et~al. 2004),\footnote{There is an ongoing debate regarding the
  bi-modality of the radio-loudness distribution of quasars (e.g.,
  Kellermann et~al. 1989; Ivezi\'{c} et~al. 2002, 2004; Cirasuolo
  et~al. 2003; Singal et~al 2011, 2012). Some recent studies with new
  approaches, e.g., investigating the loci of quasars in the
  optical/UV spectroscopic parameter space (Zamfir et~al. 2008) or the
  quasar radio luminosity function (Kimball et~al. 2011b) found two
  distinct populations with different radio properties, supporting the
  dichotomy of radio-loud vs. radio-quiet quasars (but also see, e.g.,
  Broderick \& Fender 2011).} which was defined by Kellermann
et~al. (1989) as $R=f_{\rm 5\;GHz}/f_{4400\mbox{\rm~\scriptsize\AA}}$
(rest-frame; see \S\ref{xray} for a detailed definition). Radio-quiet
quasars (RQQs) have $R<10$, while radio-loud quasars (RLQs) are those
with $R\geqslant10$. 

The basic \xray\ properties of \zfour\ RQQs are now well established
(e.g., Kaspi et~al. 2000; Brandt et~al. 2001, 2002; Vignali
et~al. 2001, 2003a, 2003b, 2005; Shemmer et~al. 2005, 2006). The
\zfour\ RQQs have remarkably similar \xray\ properties to those of
RQQs at lower redshift, which is notable given the strong evolution of
the quasar population over cosmological timescales (e.g., Richards
et~al. 2006b; Croom et~al. 2009; Jiang et~al. 2009). These results
suggest the mode of SMBH growth via accretion does not strongly evolve
with redshift.  

RLQs feature powerful jets that originate near the SMBH
and may extend up to hundreds of kpc. These jets possess bulk relativistic
velocities on at least sub-kpc scales and generate strong radio emission
that is particularly pronounced in low-inclination (highly beamed,
foreshortened, core-dominated) RLQs. RLQs usually show an enhancement of
nuclear \xray\ emission compared to RQQs, likely due to a
contribution from the spatially unresolved jet (e.g., Worrall et al.~1987;
Miller et al.~2011; and references therein). Previous \xray\ studies
of representative RLQs at $z \simgt 4$ ($R \approx 40$--$400$; e.g., Bassett
et al.~2004; Lopez et al.~2006; Saez et al.~2011) have shown that the
factor of \xray\ emission enhancement of these objects is similar to
that of low-redshift RLQs of similar radio loudness. These results limit
any redshift dependence of the nuclear \xray\ emission in RLQs
(consistent with earlier studies to moderate redshifts; e.g., Worrall et
al.~1987). The lack of strong redshift evolution argues against a dominant
\xray\ generation mechanism for the nuclear jet-linked contribution
whereby relativistic jet electrons upscatter cosmic microwave background
photons into the \xray\ band (the IC/CMB model; e.g., Tavecchio et
al.~2000; Celotti et al.~2001). While photon fields associated with the
quasar are expected to be more relevant on small scales ($\approx$~0.1--2~kpc), for $z=4$
core-dominated RLQs (for which $1^{\prime\prime}$ is 7.1 projected kpc and typical
inclinations of $5$--$7^{\circ}$ correspond to deprojected lengths $\sim$10
times greater) the spatially unresolved \xray\ emission potentially
includes contributions from the jet to tens of kpc. The IC/CMB model is
often proposed to explain the \xray\ knot emission from low-redshift 
large-scale RLQ jets studied with {\it Chandra\/}, but here too the
consequent predicted strong increase with redshift in the \xray\
luminosities of jet knots (absolute or relative to jet radio or non-IC/CMB
core \xray\ emission; e.g., Schwartz 2002) is to date not observed
(e.g., Bassett et al.~2004; Lopez et al.~2006; Marshall et al.~2011; Saez
et al.~2011). While some degree of IC/CMB emission 
is required, the relative contribution (and associated spatial scales) to
the observed \xray\ emission is currently unclear. 

Some \zfour\ quasars with the highest values of radio loudness ($R\gtrsim400$) have
been studied individually in \hbox{X-rays} (e.g., PMN~J0525$-$3343, Worsley et~al. 
2004a; Q~0906$+$6930, Romani 2006; RX~J1028.6$-$0844, Yuan et~al. 2005; GB~1428$+$4217, 
Worsley et~al. 2004b; GB~1508$+$5714, Siemiginowska et~al. 2003, Yuan et~al. 2003, 2006). 
These objects are often referred to as ``blazars'' due to their extreme radio 
loudnesses and other properties.\footnote{The term ``blazar'',
  including subcategories of BL~Lac objects and optically violently
  variable quasars (e.g., see \S1.3 of Krolik 1999), has been defined
  based on a set of criteria, including the equivalent widths of
  emission lines, the non-thermal jet-linked  
contribution to the optical continuum, broad-band variability, the radio spectral 
shape, and the polarization. However, this terminology has sometimes been applied either 
inconsistently or based on incomplete information. For our objects, it
is often not feasible to classify them clearly as ``blazars'' based on
available data. All of them are broad-line quasars (i.e., not BL~Lac
objects).  
All but one (J0913$+$5919) of our objects have flat radio spectra (see \S\ref{xray}). In this
work, we have made minimal use of the term ``blazar''. Instead we define 
a highly radio-loud quasar sample based on radio loudness (see \S\ref{sample} 
for details).} Most of the previous studies were focused on their \xray\ spectral properties. 
These objects often show soft \xray\ spectral flattening; the fraction of objects with such flattening 
appears to rise with redshift (e.g., Fiore et~al. 1998; Page et~al. 2005; Yuan et~al. 2006). 
This flattening has been proposed to be caused by intrinsic \xray\ absorption with a column density of 
\hbox{$N_H\sim(1$--$3)\times10^{22}$~cm$^{-2}$} (but also see Behar et~al. 2011). This level of intrinsic 
\xray\ absorption is also found 
in one \zfour\ quasar with moderate radio loudness (SDSS~J0011$+$1446;
Saez et~al. 2011). An alternative mechanism for the soft \xray\
spectral flattening is the bulk Comptonization of broad-line
photons by relativistic plasma moving along the RLQ jet (e.g.,
Begelman \& Sikora 1987; Sikora et~al. 1997). Celotti et~al. (2007) studied the time-dependent
spectra generated by this mechanism and reproduced the soft \xray\
flattening feature in the \xmm\ spectrum of GB~1428$+$4217. Volonteri 
et~al. (2011) studied hard-X-ray-selected high-redshift blazars and found a deficit of RLQs 
at $z>3$ compared to the expectations from the number of blazars. Possible explanations of this 
deficit require either a decrease of average bulk Lorentz factor in relativistic jets or an increase 
by a factor of 2--10 in the number of very massive black holes ($M_{\rm BH}>10^{9}M_\odot$) 
at high redshifts (derived from the radio luminosity function). They stressed the 
importance of finding high-redshift blazars which could provide constraints on the 
\hbox{SMBH-dark halo} connection and structure formation. 

Although these highly radio-loud quasars (HRLQs hereafter) are rare, they produce among the 
most powerful relativistic jets in the Universe. These remarkable objects have required the 
larger sky coverage of modern wide-field optical/UV and radio surveys to be discovered in 
sufficient numbers to assemble a statistically meaningful sample at high redshift. 
Current major \xray\ missions provide coverage of the 2--40~keV energy
band in the rest frame of \zfour\ HRLQs.  
Studies of the \xray\ and broad-band properties of high-redshift HRLQs
provide probes of the jet-launching and  
radiation mechanisms in the early stages of cosmic evolution (e.g., Fabian et~al. 1998, 1999). 
However, there 
has not been a systematic survey to investigate the general properties of these remarkable 
HRLQs at \zfour. Miller et~al. (2011; M11 hereafter) studied the \xray\ properties of a large, 
diverse sample of RLQs\footnote{M11 defined ``RLQs'' as quasars 
with $R\geqslant100$, while objects with $10\leqslant R<100$ were classified as 
radio-intermediate quasars (RIQs). To simplify the terminology, we refer to 
all quasars with $R\geqslant10$ as RLQs in this work.} with wide ranges of redshift 
and radio loudness. However, HRLQs at \zfour\ only made up $\lesssim1\%$ of 
their full sample. In this work, we conduct the first \xray\ survey of a sample 
of HRLQs at \zfour\ using new \xray\ observations by \chandra\ and sensitive 
archival \xray\ data. Our science goals include the following: (1)
assess any \xray\ emission enhancement of these objects and compare
them to HRLQs at lower redshift; (2) provide constraints on the basic
\xray\  
spectral properties of the newly observed \chandra\ targets, and study the relation between 
\xray\ spectral properties and radio loudness; (3) investigate relations
between relative \xray\ brightness, radio loudness, and optical/UV emission-line properties; 
and (4) study the broad-band spectral energy distributions (SEDs) of our objects.

In \S\ref{sample} we describe the selection of our sample of HRLQs 
at \zfour. In \S\ref{xray} we detail the \xray\ observations and 
the data analyses. Overall results and discussion are presented in \S\ref{discuss}. 
Throughout this paper, we adopt a cosmology with
$H_0=70.5$~km~s$^{-1}$~Mpc$^{-1}$, 
$\Omega_{\rm M}=0.274$, and 
$\Omega_{\Lambda}=0.726$
(e.g., Komatsu et~al. 2009). 


\section{Sample Selection}\label{sample}

\begin{figure*}[t]
    \centering
    \includegraphics[width=5.8in]{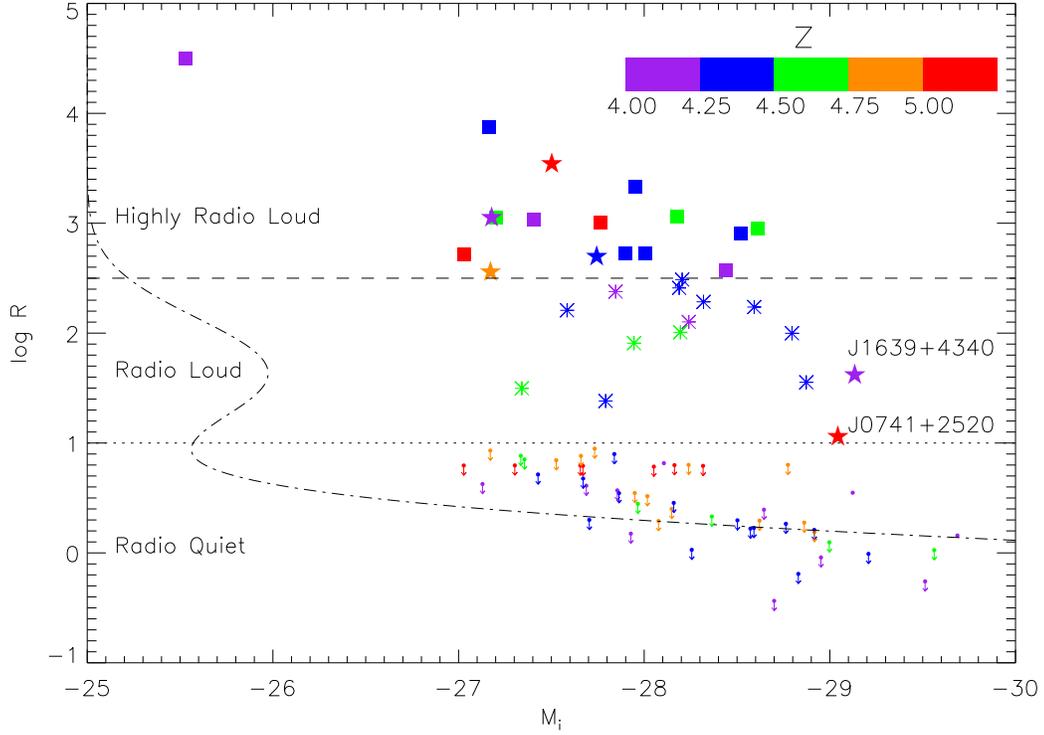}
    \caption{\footnotesize{The distribution of our sample of RLQs in the plane of $M_i$ (SDSS absolute $i$-band magnitude) vs. $\log R$ (the logarithm of 
              radio loudness), compared to high-redshift, moderately radio-loud quasars and RQQs. The filled
              stars show our \chandra\ Cycle 12 targets, while the filled squares are HRLQs at \zfour\ with 
              sensitive archival \xray\ coverage. The asterisks show the high-redshift, moderately radio-loud quasars 
              reported in Bassett et~al. (2004), Lopez et~al. (2006), and M11. The small dots represent the high-redshift, 
              radio-quiet SDSS quasars that have sensitive \xray\ coverage. All the 
              symbols are color-coded based on their redshifts using the color bar at the top right corner of the
              figure. The dotted and dashed lines show our criteria for RLQs ($\log R\geqslant1$) and highly 
              radio-loud quasars ($\log R>2.5$). The dash-dotted curve shows the quasar radio-loudness distribution 
              from Ivezi\'{c} et~al. (2004), which illustrates that our sample represents the quasars residing in the tail of high radio loudness.}
             \label{MiR_fig}}
\end{figure*}

We began our sample selection from the SDSS Data Release 7 (DR7; 
Abazajian et~al. 2009) quasar catalog (Schneider et~al. 2010) which 
covers 9380~deg$^2$ of sky area. We searched for $z\gtrsim4$ RLQs 
satisfying one or both of the following criteria: (1) $\log R > 2.5$ ($R\gtrsim320$); 
(2) $M_i < -29$. Seven objects were first selected via this method. 
We also added another object (J0741$+$2520) which does not have SDSS spectroscopy, but was confirmed as a 
very luminous ($M_i=-29.04$) $z=5.194$ RLQ in follow-up optical/UV spectroscopy of 
radio-selected SDSS sources (McGreer et~al. 2009). Among the SDSS-covered $z\gtrsim4$ RLQs without sensitive 
\xray\ coverage, these eight objects have the most remarkable radio and/or optical properties. They 
are ideal targets, with high radio loudness, high optical luminosity, and high redshift, for 
economical \chandra\ snapshot observations. 
Six of the eight objects were awarded \chandra\ time in Cycle~12 (see Table~\ref{log_table} 
for an \xray\ observation log). Four of the six targets (J1026$+$2542, J1412$+$0624, J1420$+$1205, 
and J1659$+$2101) have radio loudness \hbox{$\log R > 2.5$}. We define this radio-loudness value as 
the threshold for being ``highly radio loud''. RLQs satisfying this criterion represent the top 
$\sim5\%$ of the RLQ population in terms of radio loudness (see the dash-dotted curve 
in Fig.~\ref{MiR_fig}). We note that Sbarrato et~al. (2012b) have also
recently studied J1026$+$2542, claiming that it is a blazar. The other
two targets (J0741$+$2520 and J1639$+$4340) are only
moderately radio loud (with $\log R=1.06$ and $\log R=1.62$, respectively), but they are the most 
optically luminous ($M_i < -29$) RLQs at $z\gtrsim4$ that have sensitive \xray\ 
coverage (see Fig.~\ref{MiR_fig}). 

In addition to our \chandra\ Cycle~12 targets, we further searched for other HRLQs at 
\zfour\ satisfying our $\log R$ criterion in the sky 
area north of $\delta=-40^{\circ}$ using the NASA/IPAC Extragalactic Database 
(NED).\footnote{http://ned.ipac.caltech.edu/.} The sky area north of $\delta=-40^{\circ}$ 
is covered by the 1.4~GHz NVSS survey, and thus we are able to calculate the radio loudness for each 
object. In fact, for typical \zfour\ RLQs identified in current wide-field surveys ($m_i \lesssim 21$), 
if an object satisfies our $\log R > 2.5$ criterion for being a HRLQ, it should have been detected 
by NVSS according to 
the NVSS sensitivity \hbox{($\sim2.5$~mJy)}. Therefore, our selection method should not introduce incompleteness 
owing to the radio-flux limit.\footnote{However, it is still possible that some relevant HRLQs at \zfour\ 
have not been identified due to the lack of optical/UV spectroscopic follow up. Therefore it is not 
possible to generate a fully complete sample of \zfour\ HRLQs at present with available databases.}
For objects with both FIRST and NVSS detections, we adopt 
the radio-flux values from the FIRST catalog due to its better sensitivity; the radio flux 
values in the FIRST and NVSS catalogs are generally consistent with each other. 
A total of 24 HRLQs at \zfour\ were identified. We checked for sensitive archival \xray\ coverage 
for these \zfour\ HRLQs by \chandra, \xmm, \rosat,\footnote{We required the \rosat\ observations to 
be pointed observations (i.e., not \rosat\ All-Sky Survey) with an exposure time greater than 5~ks 
and an off-axis angle less than $19^\prime$ (i.e., within the inner ring of the PSPC detector) to achieve 
adequate sensitivity. However, none of the \zfour\ HRLQs is covered by \rosat\ observations satisfying 
these criteria.} or \swift. 
Thirteen have sensitive archival \xray\ coverage 
(see Table~\ref{log_table}),\footnote{Only one known \zfour\ HRLQ (PKS B1251$-$407) with sensitive 
\xray\ coverage lies south of $\delta=-40^{\circ}$, and thus is not included in our sample. See Yuan 
et~al. (2006) for the \xray\ properties of this object.} while the other 11 do not (see 
Table~\ref{noxray_table}).\footnote{The NED 
lists SDSS~J112429.62$+$283125.8 as a $z=4.38$ quasar based on SDSS DR6 data. 
However, the redshift of this object provided in the SDSS DR7 quasar catalog is $z=1.36$. 
Therefore we do not consider this object as a \zfour\ HRLQ although it satisfies our $\log R$ criterion. 
This object does not have sensitive \xray\ coverage.}  
Table~\ref{log_table} and Table~\ref{noxray_table} together provide a complete list of known 
HRLQs at \zfour\ in the sky 
area north of $\delta=-40^{\circ}$. All of the 13 objects having sensitive archival \xray\ coverage 
are detected in \hbox{X-rays}; all of them were targeted in their relevant \xray\ observations except for 
SDSS~J1235$-$0003. The archival \chandra, \xmm, or \swift\ observations of four objects 
(PMN~J1155$-$3107, SDSS~J1235$-$0003, GB~1713$+$2148, and PMN~J1951$+$0134) are reported in 
this work for the first time. Some of the archival objects have been classed as 
``blazars'' and were covered by multiple \xray\ observations (see \S\ref{intro}). They were usually 
selected from radio (e.g., Green-Bank 6~cm, GB6, Gregory et~al. 1996; the Parkes-MIT-NRAO survey, PMN, Griffith \& Wright 1993)
or \xray\ (e.g., the \rosat\ All Sky Survey, Voges et~al. 1999) surveys. The individual 
\xray\ properties of these blazars have been studied in detail (see the last 
column of Table~\ref{log_table} for references). In this work, we combine 
the 13 archival sources and the 4 newly observed HRLQs to form a sample to perform general 
systematic studies of HRLQs at high redshift. Our sample of HRLQs at \zfour\ with sensitive 
\xray\ coverage thus includes 17 objects, eight of which have their \xray\ properties presented 
for the first time in this work. 

We searched for high-resolution radio images of these objects obtained by the Very Long Baseline 
Array (VLBA) or Very Long Baseline Interferometry (VLBI); seven of them have available 
high-resolution images. Two objects have point-like source profiles and no indications of 
extended radio emission (J0913$+$5919, Momjian et~al. 2004;
J1235$-$0003, Momjian et~al. 2004). 
Another three objects have clear 
extended radio emission on milliarcsecond (mas) scales from relativistic jets although their 
extended radio emission only contributes $<10\%$ of the total radio flux (Q~0906$+$6930, Romani 
et~al. 2004; GB~1508$+$5714, Cheung 2004, Helmboldt et~al. 2007;
GB~1428$+$4217, Cheung et~al. 2012). The extended radio emission of GB~1508$+$5714 
coincides with the extended \xray\ emission, showing the signature of
a relativistic jet (Cheung 2004). The extended radio feature of
GB~1428$+$4217 in its VLA imaging at observed-frame 1.4~GHz and 4.9~GHz (Cheung et~al. 2012) lies
$3.6^{\prime\prime}$ away from the core. It also coincides with the
extended \xray\ emission found in its \chandra\
observation. GB~1428$+$4217 furthermore shows mas-scale extended radio
emission in its VLBI images at observed-frame 2.3~GHz and 8.6~GHz.
The extended radio emission of J1026$+$2542 is on a $\approx20$ mas scale; it contributes $\approx40\%$ 
of the total radio flux (Helmboldt et~al. 2007). The mas-scale extended radio emission of the other object, J1659$+$2101, 
contributes $\approx30\%$ of the total 
radio flux at observed-frame 1.6~GHz, while the image of this object at observed-frame 5~GHz shows 
no evidence of extended radio emission (Frey et~al. 2010).\footnote{We also found high-resolution 
radio images for two \zfour\ HRLQs without sensitive \xray\ coverage listed in Table~\ref{noxray_table}. 
The VLBI image of J0813$+$3508 shows jet-like resolved structure at observed-frame 1.6~GHz, which 
contributes $\approx1/4$ of the total flux, while its image at observed-frame 5~GHz does not have 
evidence of extended emission (Frey et~al. 2010). J1242$+$5422 has extended radio emission in both of 
its VLBI images at observed-frame 1.6~GHz and 5~GHz, where the extended emission only contributes $\approx3\%$ 
of the total fluxes (Frey et~al. 2010).} The five objects with extended radio emission 
have higher radio loudness on average ($\langle\log R\rangle = 3.21\pm0.23$) than the 
three objects without extended radio emission ($\langle\log R\rangle = 2.81\pm0.25$).

Besides the 17 HRLQs at \zfour, we will also separately discuss the two highly optically luminous 
RLQs observed by \chandra\ in Cycle~12 even though they do not satisfy the ``highly radio loud'' 
criterion. All of our 19 objects are among the most-luminous RLQs 
in both the radio and optical/UV bands (see Fig.~\ref{zluv_fig}). The monochromatic luminosity 
ranges of our objects are \hbox{$10^{33}$--$10^{36}$~erg~s$^{-1}$~Hz$^{-1}$} in the radio 
(at rest-frame 5~GHz) and \hbox{$10^{30}$--$10^{33}$~erg~s$^{-1}$~Hz$^{-1}$} in the optical/UV 
(at rest-frame 2500~\AA). GB~1713$+$2148 has remarkable properties even compared to other 
HRLQs in our sample. It is much fainter (by a factor of $\approx6$) in the optical band, 
and it has the highest radio loudness (see Fig.~\ref{MiR_fig}). Its radio loudness is also 
higher than those of most of the objects ($99\%$) in the full sample of M11.
All of our objects are spectroscopically confirmed 
broad-line quasars by either the SDSS or other observations. They have precise redshift 
measurements based on their broad emission lines (e.g., Ly$\alpha$ and/or C~{\sc iv}). 
The available SDSS spectra of nine objects (five \chandra\ Cycle~12 targets and 
four archival sources) are shown in Fig.~\ref{allspec_fig}. 


\section{X-ray Data Analysis}\label{xray}

The six RLQs at \zfour\ targeted by \chandra\ in
Cycle~12 were observed with the S3 CCD of the Advanced
CCD Imaging Spectrometer (ACIS; Garmire et~al. 2003). The
\chandra\ data were reduced using standard CIAO v4.3 routines. 
We generated \xray\ images for the observed-frame
soft \hbox{(0.5--2.0 keV)}, hard \hbox{(2.0--8.0 keV)}, and full
\hbox{(0.5--8.0 keV)} bands using {\it ASCA} grade 0, 2, 3, 4, and
6 events. The source detection was performed using the {\sc wavdetect} 
algorithm (Freeman et~al. 2002) with a detection threshold of $10^{-6}$ and
wavelet scales of $1$, $\sqrt{2}$, $2$, $2\sqrt{2}$, and $4$
pixels. All targets were clearly detected by \chandra\
within $0.6''$ of the optical coordinates. To assess possible extended \xray\ 
emission from a putative jet, we followed the method in \S2.3 of Bassett et~al. (2004) 
requiring a minimum of $\approx3$ nearby counts offset from the core by 
$\approx2^{\prime\prime}$ along roughly the same direction. None of our targets shows evidence for 
extended \xray\ emission. The \xray\ counts were 
measured with aperture photometry using the IDL {\sc aper} procedure. 
The aperture radius was adopted as $1.5''$ for each source
($\approx95\%$ enclosed energy for the soft band, $\approx90\%$
enclosed energy for the hard band; aperture corrections were applied). 
The background counts were retrieved from an annular region with inner 
and outer radii of twice and three times the aperture radius, and 
were scaled to the source-aperture area. All background
regions are free of \xray\ sources. Table~\ref{cts_table} lists the
\xray\ counts in the three bands defined above, as well as the band ratio
(defined as the ratio between hard-band and soft-band
counts) and effective \pl\ photon index for each source. The
effective photon index was derived from the band ratio
with the \chandra\ PIMMS\footnote{http://cxc.harvard.edu/toolkit/pimms.jsp} tool,
under the assumption of a \pl\ model with only Galactic absorption.

The \xray\ properties of the archival objects were obtained from the literature (see 
Table~\ref{log_table}) except for those objects listed in Table~\ref{cts_table} for which 
their archival \chandra, \xmm, or \swift\ observations have not been previously published. 
For objects with multiple \xray\ observations, we chose the \chandra/\xmm\ observation with 
the longest exposure time.\footnote{We did not use the \xmm\ observation with the longest 
exposure time for GB~1428$+$4217 because this object was undergoing an exceptional 
radio flare (see Worsley et~al. 2006).} This approach should not introduce any biases in the 
following considerations regarding the \xray\ emission strength. Vignali et~al. (2003a) reported a tentative 
detection of GB~1713$+$2148 using a \rosat\ HRI observation. This object was later observed 
with \chandra\ ACIS in Cycle~5 (PI: L. Gurvits). We processed 
the \chandra\ data for GB~1713$+$2148 using the same method as described in the previous 
paragraph. The \xray\ flux of GB~1713$+$2148 in the \chandra\ epoch is consistent with 
that in the \rosat\ epoch, showing that this object did not have any strong \xray\ variability. 

SDSS~J1235$-$0003 was serendipitously covered by an \xmm\ observation on 2010 July 01. We performed 
data reduction and processing with standard \xmm\ Science Analysis System (SAS; v10.0.0) routines. 
\xray\ images were generated for the observed-frame soft, hard, and full bands for the \verb+MOS1+ and 
\verb+MOS2+ detectors; this object is not covered by the \verb+pn+ detector. 
Source detection was carried out using the {\sc eboxdetect} procedure. This 
object was detected only in the \verb+MOS1+ full-band image at a $\approx3\sigma$ level,
and thus we only used the \verb+MOS1+ image for 
its \xray\ photometry (it was not detected in the \verb+MOS2+ images because it is very close to the 
CCD edge). We filtered the events file by removing the periods of background flaring (49\% of 
the total exposure time). The aperture radius for the photometry ($53.0^{\prime\prime}$) was the 90\% enclosed-energy 
radius at 1.5~keV based on the point spread function (PSF) of the \verb+MOS1+ detector at an off-axis angle 
of $12.8^\prime$.  The upper limits on \xray\ counts in the soft and hard bands were calculated as $3\sqrt{N}$, 
where $N$ is the total counts in the aperture. The vignetting was corrected by calculating the effective 
exposure time (4.2~ks) at the source location based on the exposure map. 

Another two archival objects, PMN~J1155$-$3107 and PMN~J1951$+$0134, were targeted by the \swift\ X-ray Telescope (XRT; Burrows et~al. 2005) in two 
(ObsIDs: 00036263001/2, total exposure time 5.1~ks) and three (ObsIDs: 00036791001/2/3, total exposure time 10.4~ks) observations, respectively.
For each object, we first generated the exposure maps for all XRT events files using the {\sc xrtexpomap} routine. 
We then merged the event files and exposure maps for each object using the {\sc xselect} and {\sc ximage} packages, 
respectively. Similar to the approach for the \chandra\ and \xmm\ data, XRT images were created for the observed-frame soft, hard, 
and full bands. Source detection was performed with the {\sc wavdetect} algorithm with a detection threshold of $10^{-6}$. 
PMN~J1155$-$3107 is detected in the full band and soft band, while PMN~J1951$+$0134 is detected in all three bands. 
Aperture photometry was performed using aperture 
radii of $59.6^{\prime\prime}$ and $64.9^{\prime\prime}$, which are the $90\%$ enclosed-energy radii in the full band based on 
the PSF of the XRT detector at off-axis angles of $1.0^\prime$ and $2.6^\prime$,\footnote{The \swift\ mission has 
excellent rapid pointing capability, while its pointing accuracy is not exceptional. Therefore, it is normal to have off-axis angles of 
$\approx1^\prime$--2$^\prime$ for targeted observations.} respectively. The \xray\ photometry of these two archival objects is also 
listed in Table~\ref{cts_table}.

Table~\ref{aox_table} lists the key \xray, optical, and radio properties of our sample:

\noindent Column (1): The name of the quasar.



\noindent Column (2): The apparent SDSS $i$-band magnitude of the quasar. For objects 
covered by the SDSS footprint, the values are obtained either from the SDSS DR7 quasar 
catalog (BEST photometry; Schneider et~al. 2010) or from the SDSS database. For other 
objects, the $i$-band magnitude was converted from AB$_{1450(1+z)}$ under the assumption 
of a \pl\ spectral index of $\alpha_\nu\;=\;-0.5$ ($f_\nu\propto\nu^{\alpha_\nu}$; e.g., Vanden Berk 
et~al. 2001). The values of 
AB$_{1450(1+z)}$ were calculated from $R$-band magnitudes (from the NED) using the empirical relation of 
$AB_{1450(1+z)}= R-0.684z+3.10$ (Kaspi et~al. 2000; Vignali et~al. 2003a). 

\noindent Column (3): The absolute SDSS $i$-band magnitude for the quasar, $M_{i}$, 
from the SDSS DR7 quasar catalog, which was calculated 
from the apparent SDSS $i$-band magnitude in Column~2 by correcting for 
Galactic extinction and assuming a \pl\ spectral index of $\alpha_\nu=-0.5$.

\noindent Column (4): The Galactic neutral hydrogen column density calculated  
with the \chandra\ COLDEN\footnote{http://cxc.harvard.edu/toolkit/colden.jsp} 
tool (Dickey \& Lockman 1990; Stark et~al. 1992), in units of $10^{20}$~cm$^{-2}$.

\noindent Column (5): The count rate in the observed-frame soft \xray\ band 
($0.5$--$2.0$~keV) for the \chandra\ observed objects, in units of $10^{-3}$~s$^{-1}$. 

\noindent Column (6): The \xray\ flux in the 
observed-frame soft band (\hbox{$0.5$--$2.0$~keV}) corrected for Galactic 
absorption. This measurement was obtained with the 
\chandra\ PIMMS tool and is in units of $10^{-14}$~erg~cm$^{-2}$~s$^{-1}$. An 
absorbed \pl\ model was used with the photon index ($\Gamma_{\rm X}$) listed in Column~9 and 
the Galactic neutral hydrogen column density ($N_H$) listed in Column~4. For objects 
without $\Gamma_{\rm X}$ information or only with lower limits upon $\Gamma_{\rm X}$, we adopt $\Gamma_{\rm X}=1.6$ 
which is typical for radio-loud quasars (e.g., Page et~al. 2005).

\noindent Column (7): The \xray\ flux density at rest-frame 2~keV generally obtained from the count rate 
in the observed-frame 0.5--2.0~keV band with PIMMS, in units of $10^{-32}$ erg cm$^{-2}$~s$^{-1}$~Hz$^{-1}$ and 
corrected for Galactic absorption. Although for our objects rest-frame 2~keV corresponds to observed-frame 
0.3--0.4~keV, we chose the standard approach to using the observed-frame 0.5--2.0~keV band because we are able to 
minimize effects from potential intrinsic \xray\ absorption and \xray\ instrumental contamination below 
observed-frame 0.5~keV, and we can also obtain better photon statistics with the larger numbers 
of \xray\ counts in this band.

\noindent Column (8): The logarithm of the quasar luminosity (erg s$^{-1}$) in the rest-frame 2--10~keV band 
corrected for Galactic absorption.

\noindent Column (9): The power-law photon index ($\Gamma_{\rm X}$) of the X-ray spectrum. For 
objects listed in Table~\ref{cts_table}, the values were obtained 
from band-ratio analysis (see the last column of Table~\ref{cts_table}). 
For the other objects, the values were retrieved from the literature. Some objects have 
shown \xray\ spectral variability (e.g., GB~1428+4217; see Fig.~4 of Worsley et~al. 2006). 
The choice of $\Gamma_{\rm X}$ values do not significantly affect the \xray\ flux calculation. 
We have chosen the \chandra/\xmm\ observations with the longest exposure time which should 
best constrain the power-law photon index. 

\noindent Column (10): The continuum flux density at rest-frame 2500~\AA~ in 
units of $10^{-27}$ erg~cm$^{-2}$~s$^{-1}$~Hz$^{-1}$. For objects appearing in the SDSS DR7 quasar 
catalog, the values were obtained from the catalog of Shen et~al. (2011). For other SDSS covered objects, 
the flux density was calculated using the SDSS photometry with the composite spectrum for SDSS quasars 
in Vanden Berk et~al. (2001). For objects not covered by the SDSS, the flux density was calculated using 
the optical photometry from the literature under the assumption of a \pl\ spectral index of $\alpha_\nu\;=\;-0.5$.

\noindent Column (11): The logarithm of the 
monochromatic luminosity (erg s$^{-1}$ Hz$^{-1}$) at rest-frame 2500~\AA, which was calculated from the flux density at 
rest-frame 2500~\AA\ given in Column~9. A standard cosmological bandpass correction was applied when converting flux into 
monochromatic luminosity.

\noindent Column (12): The radio \pl\ slope $\alpha_{\rm r}$ ($f_\nu\propto\nu^{\alpha_{\rm r}}$) 
between observed-frame 1.4~GHz and 5~GHz. The 1.4~GHz flux density was obtained from the FIRST 
or NVSS surveys, while the 5~GHz flux density was obtained from the GB6 survey, the PMN survey, 
or from the literature for individual objects (J1639$+$4340, Holt
et~al. 2004; Q~0906$+$6930, Romani 2006; J0913$+$5919, Momjian et~al. 2004). 
The observations at 1.4~GHz and at 5~GHz are not simultaneous. 
All objects with $\alpha_{\rm r}$ measurements are flat-spectrum radio quasars 
(FSRQs; $\alpha_{\rm r} > -0.5$) except for J0913$+$5919. For the five objects without 5~GHz
flux measurements, we assume $\alpha_{\rm r}=0$ to keep consistency
with other objects in our sample.

\noindent Column (13): The logarithm of the monochromatic luminosity (erg s$^{-1}$ Hz$^{-1}$) at rest-frame 5~GHz, obtained 
from the flux density at rest-frame 5~GHz, $f_{\rm 5~GHz}$.  This quantity was calculated using a radio \pl\ slope given in 
Column~12 and the flux density at observed-frame 1.4~GHz, $f_{\rm 1.4~GHz}$, which were obtained from 
the FIRST or the NVSS surveys. A standard cosmological bandpass correction was also applied. 

\noindent Column (14): The logarithm of the radio-loudness parameter, given by 
\begin{equation}
    \log R = \log \bigg(\frac{f_{\rm 5\;GHz}}{f_{4400\mbox{\rm~\scriptsize\AA}}}\bigg){\rm .}
\end{equation}
The denominator, $f_{4400\mbox{\rm~\scriptsize\AA}}$, was calculated from 
$f_{2500\mbox{\rm~\scriptsize\AA}}$ using an optical \pl\ slope of \hbox{$\alpha_\nu$ = $-0.5$}. 
The numerator, $f_{5\;{\rm GHz}}$, was calculated using the same method as in Column~13. 
All measures of flux density are per unit frequency. 
The definition of radio loudness in M11 used $f_{2500\mbox{\rm~\scriptsize\AA}}$ 
as the denominator. We converted their values to our definition for the scientific 
analyses described in the following section. 

\noindent Column (15): The \aox\ parameter, defined by
\begin{equation}
    \alpha_{\rm ox} = \frac{{\rm log}(f_{\rm 2\;keV} / 
    f_{2500\mbox{\rm~\scriptsize\AA}})}{{\rm log}(\nu_{\rm 2\;keV} / \nu_{2500\mbox{\rm~\scriptsize\AA}})}
    = 0.384\ {\rm log} \bigg(\frac{f_{\rm 2\;keV}}{f_{2500\mbox{\rm~\scriptsize\AA}}}\bigg),
\end{equation}
which represents the slope of an assumed power law connecting rest-frame 2500~\AA\ and 2~keV. 
Our UV and \xray\ measurements
were not simultaneous, and thus the \aox\ parameter could be affected by variability. 

\noindent Column (16): \daoxrq, defined as
\begin{equation}
    \Delta\alpha_{\rm ox, RQQ} = \alpha_{\rm ox}({\rm measured}) - \alpha_{\rm ox, RQQ}({\rm expected}).
\end{equation}
The expected $\alpha_{\rm ox, RQQ}$ value for a typical RQQ is calculated 
from the $\alpha_{\rm ox}$--$L_{2500\mbox{\rm~\scriptsize\AA}}$ correlation given as 
Equation (3) of Just et~al.~(2007). The statistical significance of this 
difference compared to RQQs (given in parentheses) is in 
units of $\sigma$, which is 
given in Table~5 of Steffen et~al.~(2006) as the RMS of \aox\ for 
several ranges of luminosity. The \daoxrq\ parameter quantifies the excess of \xray\ 
emission from the relativistic jet of the RLQs compared to that of RQQs for which the 
\xray\ emission is mainly from the accretion disk and its corona. In \S\ref{discuss:lines} 
we will validate the use of \daoxrq\ as a useful diagnostic quantity for HRLQs. 

\noindent Column (17): \daoxrl, defined as
\begin{equation}
    \Delta\alpha_{\rm ox, RLQ} = \alpha_{\rm ox}({\rm measured}) - \alpha_{\rm ox, RLQ}({\rm expected}).
\end{equation}
The expected $\alpha_{\rm ox, RLQ}$ value for a typical RLQ is obtained  
from the $L_{\rm 2~keV}$--$L_{2500\mbox{\rm~\scriptsize\AA}}$--$L_{\rm 5~GHz}$ correlation based on  
the full sample in M11 given in their Table~7, which is equivalent 
to 
\begin{equation}
    \alpha_{\rm ox, RLQ} = -0.199\ L_{2500\mbox{\rm~\scriptsize\AA}} + 0.105\ L_{\rm 5~GHz}+1.194.
\end{equation}
The \daoxrl\ parameter assesses the relative \xray\ brightness of a RLQ
compared to typical RLQs (mostly at $z=0.3$--$2.5$ and with a median redshift of $z=1.4$) 
with similar optical/UV and radio luminosities that have 
both disk/corona-linked and jet-linked \xray\ emission. 


\section{Results and Discussion}\label{discuss}

\subsection{Enhanced \xray\ Emission at High Redshift}\label{discuss:daox}

\subsubsection{The Relative X-ray Brightness of \zfour\ HRLQs}\label{discuss:daox:daox}

To assess the enhancement of the \xray\ emission of our objects at \zfour\ compared to 
similar objects at lower redshift, we compare their distributions of 
\aox, \daoxrq, and \daoxrl\ to those of RLQs at $z<4$. Fig.~\ref{raox_fig} shows the 
positions of our objects and the typical RLQs in M11 in the 
$\log R$--(\aox, \daoxrq, \daoxrl) planes. Our objects generally occupy the same region as the typical 
RLQs in the $\log R$--\aox\ plane (see the top panel of Fig.~\ref{raox_fig}). However, our sample 
appears to have an excess of objects with stronger \xray\ emission in the $\log R$--(\daoxrq, \daoxrl) planes. 
The thick black lines in the middle panel of Fig.~\ref{raox_fig} show the mean \daoxrq\ values of 
M11 objects binned by radio loudness ($\Delta\log R=0.2$ per bin). Fifteen out of our 17 HRLQs at \zfour\ have 
\daoxrq\ values greater than the mean value of the corresponding $\log R$ bin. Similarly, 14 of our 
17 HRLQs at \zfour\ have \daoxrl\ values greater than zero, which generally represents the mean \daoxrl\ value 
of M11 objects (see the bottom panel of Fig.~\ref{raox_fig}). We will show later that the generally higher 
\daoxrq\ and \daoxrl\ values of our objects are caused by stronger \xray\ emission rather than weaker optical/UV  
emission (see the broad-band SED studies on our objects in \S\ref{discuss:sed}).
It is worth noting that the \xray\ weakest object in our \zfour\ HRLQ
sample (J0913$+$5919; see the red square with the smallest \daoxrq\
and \daoxrl\ values in Fig.~\ref{raox_fig}) has a steep radio spectral
slope ($\alpha_{\rm r}=-0.67$), while the others with available
$\alpha_{\rm r}$ values are all FSRQs. 

To verify this excess quantitatively, we 
obtained the mean values and statistical distributions of \daoxrq\ and \daoxrl\ for our HRLQs
and a comparison sample consisting of objects in the full sample\footnote{The full sample of M11 contains a mixture of 
objects that are optically selected ($\sim80\%$), radio selected ($\sim15\%$), or \xray\ selected ($\sim5\%$).} of M11 
with $\log R>2.5$ and $z<4$. Our sample, which includes objects originally selected in the optical (7 objects; e.g., SDSS), radio (9 
objects; e.g., GB6 and PMN), or \xray\ (1 object; from the \rosat\ All-Sky Survey) bands, is somewhat heterogeneous. To minimize potential 
selection biases (e.g., the most exceptional objects being preferentially targeted in \xray\ observations), 
we need a sample containing most of the known \zfour\ HRLQs regardless of their selection methods 
and {\it having nearly complete \xray\ coverage}. To achieve this goal, we chose an optical-magnitude cut of $m_i<20$ on all known \zfour\ HRLQs 
north of $\delta=-40^\circ$ (see Table~\ref{log_table} and Table~\ref{noxray_table}). There are 15 HRLQs at \zfour\ 
satisfying this $m_i$ cut, and 12 of them have sensitive \xray\ coverage (and thus are included in our sample). For the 
three HRLQs without sensitive \xray\ coverage, two of them (J0813$+$3508 and J1242$+$5422, both optically selected) were in our original 
\chandra\ Cycle~12 proposal, but were not awarded \chandra\ time (see \S\ref{sample}). These two objects have the 
highest redshifts, optical brightnesses, and optical luminosities among those \zfour\ HRLQs without sensitive \xray\ coverage 
(see Table~\ref{noxray_table}). The $i$-band magnitude of the other HRLQ (J2220$+$0025, optically selected) is close to our $m_i$ cut. 
To maintain consistency, we also applied the $m_i<20$ criterion to our comparison sample from M11. There are
283 objects in total in our comparison sample, including 269 \xray\ detected objects and 14 objects with \xray\ upper limits.

The mean values of \daoxrq\ and \daoxrl\ shown in Table~\ref{twost_table} were calculated using the Kaplan-Meier 
estimator\footnote{The Kaplan-Meier estimator is applicable to censored data; note some of the M11 data points are 
censored.} implemented in the Astronomy Survival Analysis (ASURV) package (e.g., Lavalley et~al. 1992). These mean 
values indicate that HRLQs at $z>4$ have stronger \xray\ emission (by a factor of $\approx3$ on average) than those 
at $z<4$ with similar UV and/or radio luminosity.
The distributions of \daoxrq\ and \daoxrl\ are shown in Fig.~\ref{daox_fig}. The Peto-Prentice test 
(e.g., Latta 1981), also implemented in the ASURV package, was employed to determine whether the $z>4$ and $z<4$  
HRLQs follow the same \daoxrq\ and \daoxrl\ distributions. The Peto-Prentice test is preferred over 
other two-sample tests because it is the least affected by factors such as unequal sample sizes 
(e.g., Latta 1981) which do exist in our study. The distribution of \daoxrq\ for our objects is significantly 
different from that of the HRLQs at $z<4$. The null-hypothesis (i.e., the two samples following the same 
distribution) probability is only $1.21\times10^{-4}$. The two-sample test for the \daoxrl\ distribution reaches a similar 
conclusion; the null-hypothesis probability is slightly higher ($1.37\times10^{-4}$), but it still shows that the two samples 
follow different distributions at a $>3\sigma$ level. One might speculate that the significant differences of the \daox\ distributions 
are caused by a ``tail'' of \zfour\ objects with extraordinary \xray\ brightness. However, Fig.~\ref{daox_fig} suggests that the 
entire \daox\ histograms of our \zfour\ HRLQs are shifted toward high \daox\ values relative to those of $z<4$ HRLQs (also see Fig.~\ref{raox_fig}). 
Therefore, our sample of HRLQs shows a significant \xray\ emission enhancement at $z>4$ over those HRLQs at lower redshift. 
However, one should keep in mind that 
although we have compiled a sample of \zfour\ HRLQs with nearly complete \xray\ coverage using all available databases, 
potential selection biases could still exist. For example, although we have assembled all known \zfour\ HRLQs north of 
\hbox{$\delta=-40^\circ$}, this is still not a fully complete sample (see Footnote~11). 
Therefore, in the next subsection we describe a variety of tests of the robustness of our results. 

\subsubsection{Robustness Tests}\label{discuss:daox:robust}

\begin{figure*}[t]
    \centering
    \includegraphics[width=5.5in]{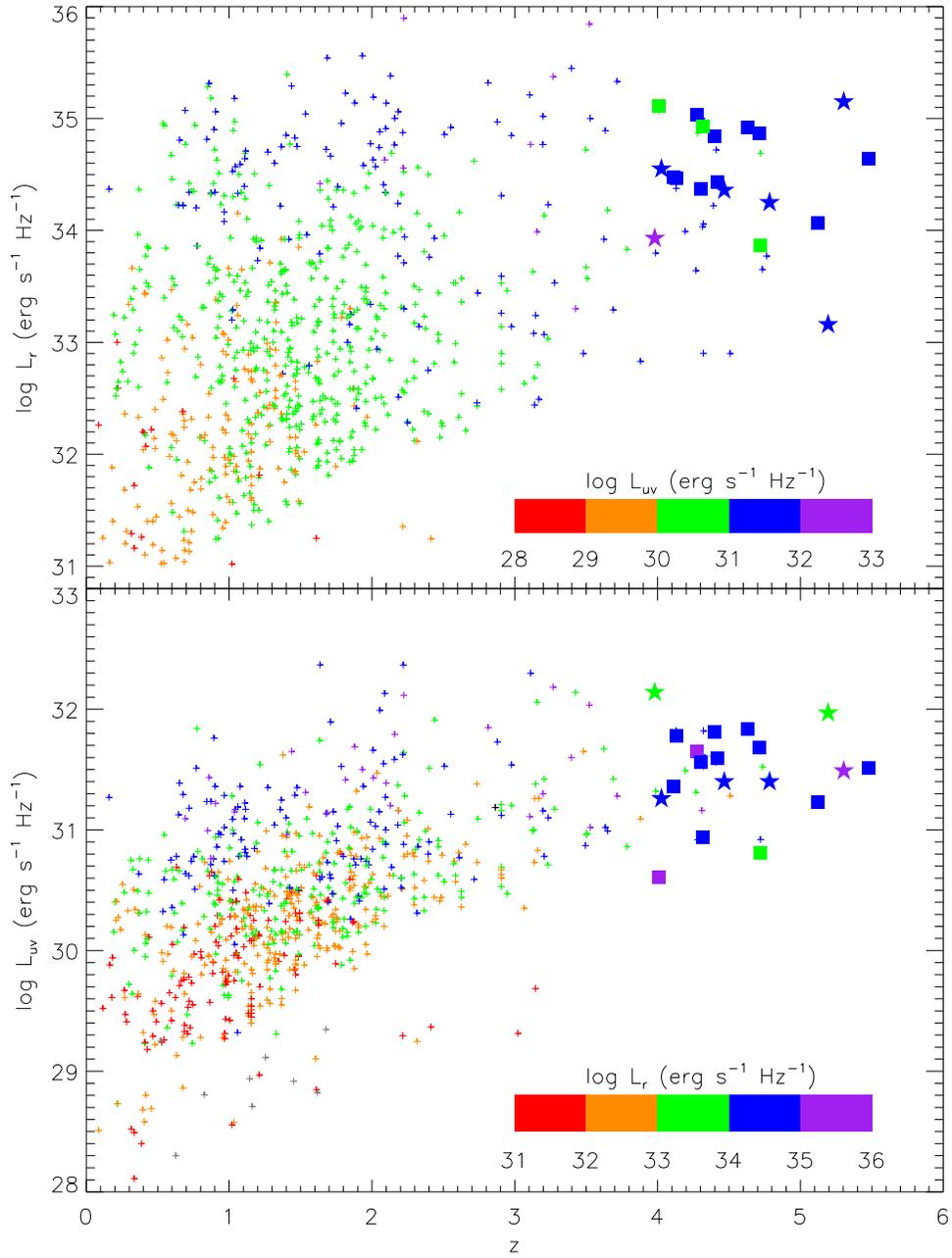}
    \caption{\footnotesize{The monochromatic luminosity at rest-frame 5~GHz (radio; upper panel) and at rest-frame 2500~\AA\ (UV; lower panel), plotted against 
              redshift. The filled
              stars show our \chandra\  Cycle 12 targets, while the filled squares are our HRLQs with 
              sensitive archival \xray\ coverage. The plus signs represent the radio-loud and radio-intermediate 
              objects in the full sample of M11. The upper (lower) panel is color-coded based on UV (radio) luminosity using the color bars at 
              the bottom right corner of each panel. Our sample of quasars are among the most luminous objects in both the radio and UV bands.} 
             \label{zluv_fig}}
\end{figure*}

\begin{figure*}[t]
    \centering
    \includegraphics[width=5.5in]{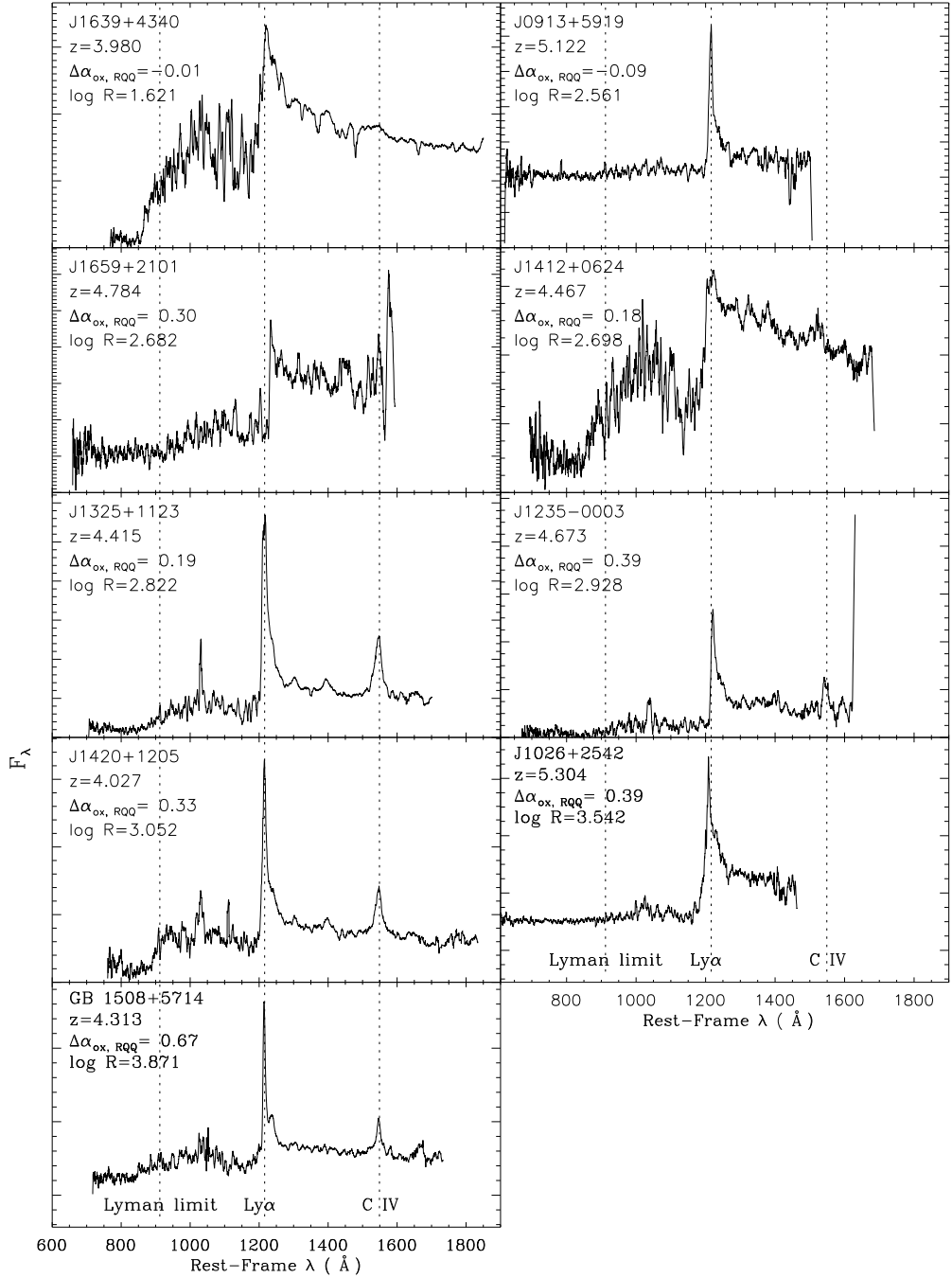}
    \caption{\footnotesize{The SDSS spectra of five of our \chandra\ Cycle 12 targets and four objects with sensitive archival \xray\ coverage, 
             ordered by $\log R$ which is shown at the top-left corner of each panel. Also labeled are the object names and redshifts. The 
             $y$-coordinates are the flux density ($F_\lambda$) in linear units. The spectra have been smoothed using a 20-pixel sliding-box filter. 
             The wavelengths corresponding to major emission lines (Ly$\alpha$~$\lambda$1216 and C~{\sc IV}~$\lambda$1549) and the Lyman limit are 
             labeled by the dotted vertical lines. The spectral resolution is $R\approx2000$. 
             The apparent broad features of J1639$+$4340 and J1412$+$0624 at $\sim1000$~\AA\ are caused by the stretched scaling 
             due to their weak Ly$\alpha$ emission lines.}
             \label{allspec_fig}}
\end{figure*}

\begin{figure*}[t]
    \centering
    \includegraphics[width=5.3in]{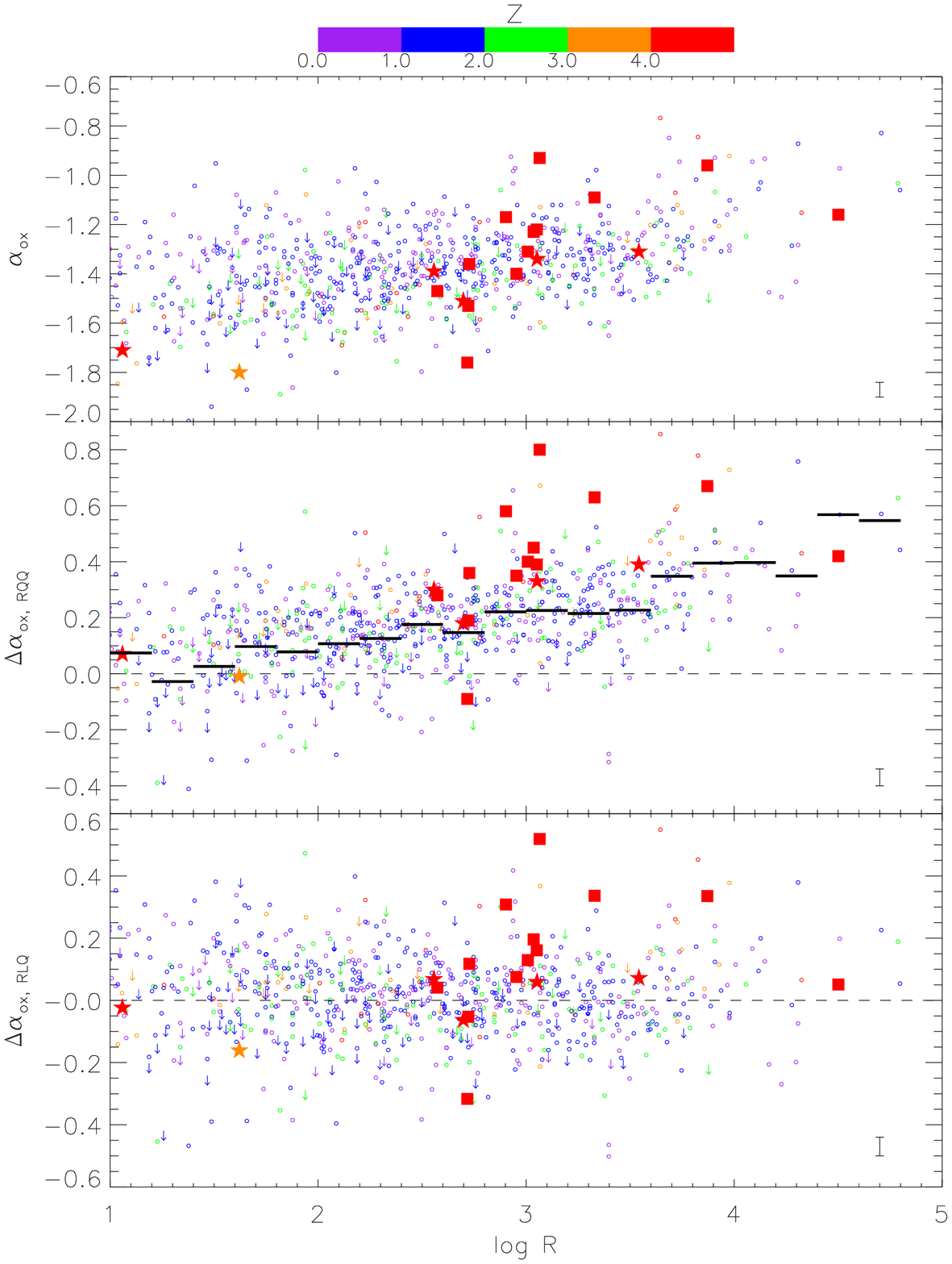}
    \caption{\footnotesize{The relation between radio loudness and \aox\ (top panel), \daoxrq\ (middle panel), and \daoxrl\ (bottom panel)
              for RLQs at all redshifts. The filled
              stars show our \chandra\  Cycle 12 targets, while the filled squares are our HRLQs with 
              sensitive archival \xray\ coverage. The open circles (downward arrows) represent the radio-loud and radio-intermediate 
              objects in the full sample of M11 that have \xray\ detections (upper limits). The typical error bars for 
              \aox, \daoxrq, and \daoxrl\ for our objects are shown at the bottom-right corner of each panel. 
              The dashed lines represent the positions of \daox$=0$. 
              The thick black lines show the mean \daoxrq\ values for the M11 objects binned in $\log R$ ($\Delta\log R=0.2$ per bin). All symbols are 
              color-coded based on their redshifts using the color bar at the top of the figure.} 
             \label{raox_fig}}
\end{figure*}

\begin{figure*}[t]
    \centering
    \includegraphics[width=5.5in]{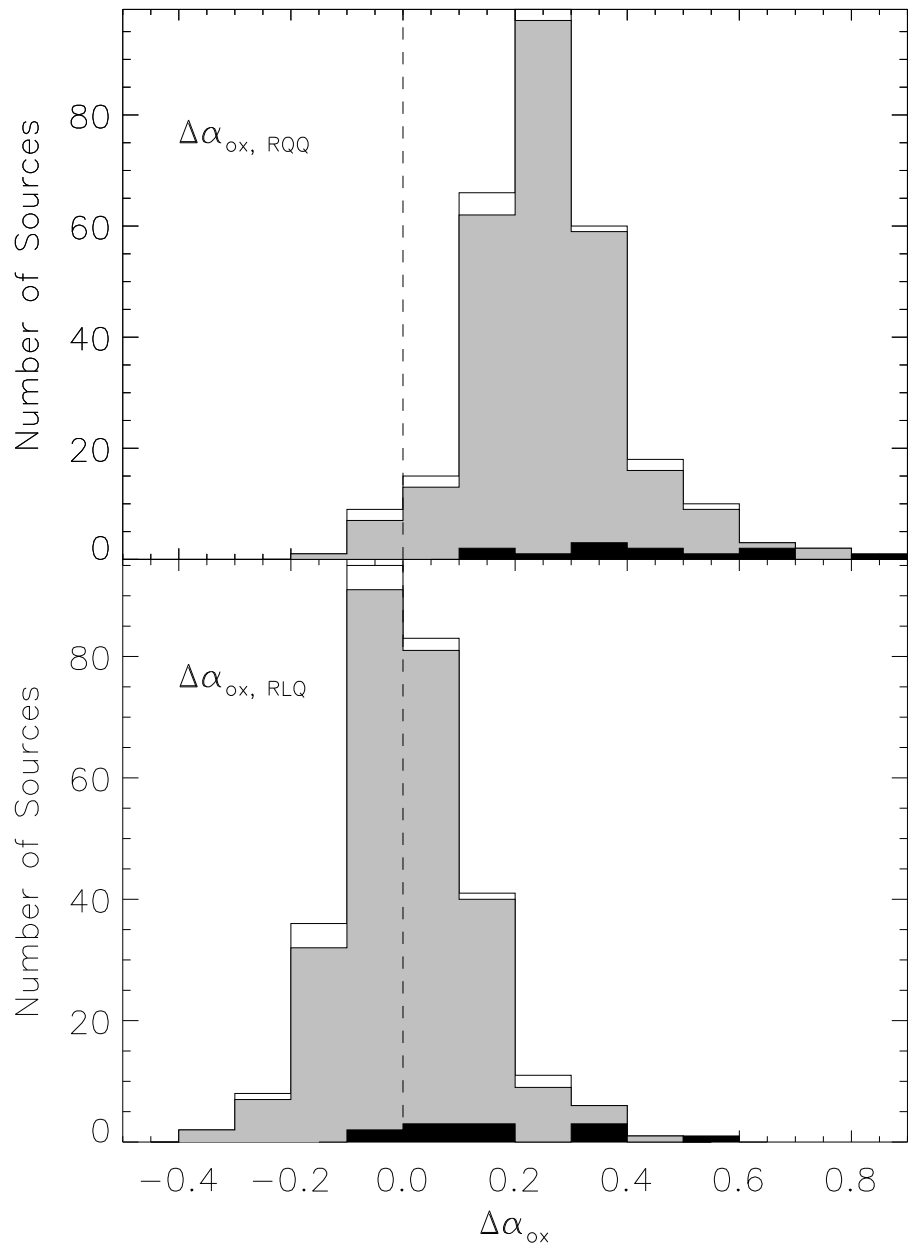}
    \caption{\footnotesize{The histograms of $\Delta\alpha_{\rm ox, RQQ}$ (top panel) and $\Delta\alpha_{\rm ox, RLQ}$ (bottom panel) for the 
             full-sample objects in M11 with $\log R>2.5$, $z<4$, and $m_i<20$ (grey and open histograms for \xray\ detected and undetected objects, 
             respectively) and our HRLQs at \zfour\ with $m_i<20$ (black histogram).}
             \label{daox_fig}}
\end{figure*}

In order to test the robustness of our results against selection issues, we carried out further two-sample analyses by 
constructing other comparison samples or controlling for a variety of relevant parameters (see results in Table~\ref{twost_table}):
\begin{enumerate}
\item Many of our \zfour\ HRLQs are radio selected, while the majority of our comparison sample from M11 is optically 
selected. To minimize the effects of different selection methods, we first compare our radio-selected \zfour\ HRLQs with a radio-selected 
subsample in M11. We chose their ``{\it Einstein}'' supplemental sample (see \S2.2.1 of M11) which was primarily radio selected. 
The \xray\ observations of these objects were first presented in Worrall et~al. (1987), and we transform their luminosities 
to our adopted cosmology. We also incorporate improved X-ray coverage for a handful of objects, as in M11. All of these objects 
have $z<4$ and $m_i < 20$;\footnote{The $m_i$ values were estimated from their monochromatic luminosities at rest-frame 
2500~\AA.} we further require them to be HRLQs (i.e., \hbox{$\log R > 2.5$}). Our \zfour\ HRLQs have significantly different \daox\ distributions 
(both \daoxrq\ and \daoxrl) from those of this radio-selected comparison sample at a $\approx3\sigma$ level.
\item We generate a quasi-radio-selected comparison sample following \S2.1.1 of M11 by only selecting those objects that were targeted 
by SDSS spectroscopy owing to being FIRST radio sources (i.e., with the ``FIRST'' flag set in the SDSS quasar catalog).\footnote{This 
is not a true radio-selected sample because, e.g., they were restricted by SDSS magnitude limits for FIRST sources. See \S2.1.1 of M11 for 
details.} Our \zfour\ HRLQs have significantly different \daox\ distributions from those of this ``radio-selected'' comparison sample at a 
$\approx4\sigma$ level. 
\item As discussed in \S\ref{sample}, our \zfour\ HRLQs are among the RLQs with the highest radio luminosities; this may introduce a 
luminosity bias into our sample. As discussed in Ghisellini
et~al. (2011), for example, more luminous blazars have relatively more hard X-ray
emission (because the peak photon energy shifts to lower frequencies), and this
could make the observed X-ray flux greater for high-redshift blazars. Therefore, we construct another comparison sample from M11 
to have comparable radio luminosity ($\log L_{\rm 5~GHz} > 34$; compare with Fig.~\ref{zluv_fig}). After controlling for radio luminosity, the significance of the difference 
in the \daox\ distributions becomes slightly lower, but still remains at a $\approx3\sigma$ level. 
\item Similarly, we construct another comparison sample to have comparable optical/UV luminosity ($\log L_{2500\mbox{\rm~\scriptsize\AA}} > 30.9$; compare with Fig.~\ref{zluv_fig}) 
to our \zfour\ HRLQs to avoid optical/UV luminosity bias. The significance of the \daox\ distribution difference is at a $\gtrsim2.5\sigma$ 
level after controlling for optical/UV luminosity.
\item All of our \zfour\ HRLQs with available $\alpha_{\rm r}$ are
  FSRQs ($\alpha_{\rm r} > -0.5$; see \S\ref{xray}) except for
  J0913$+$5919\footnote{This object is not included in our two-sample
  tests since it does not satisfy $m_i < 20$.}, which may suggest that low-inclination RLQs are 
over-represented in our sample. Therefore, for another comparison sample, we selected the M11 full-sample objects with available $\alpha_{\rm r}$ 
measurements that satisfy the FSRQ criterion. After controlling for this effect, our results still have $> 3.5\sigma$ significance.
\end{enumerate}
In summary, our finding that \zfour\ HRLQs have a substantial \xray\ emission enhancement over HRLQs at lower redshifts remains 
significant after controlling for those factors which may introduce
biases. We also note that, owing to the respectable sizes of the
samples being compared, variability effects upon \daox\ distributions
should tend to average out (also see \S3 for explanation of how our
selection of the \xray\ observations utilized was designed to avoid
biases due to variability).

It is possible that this \xray\ enhancement of HRLQs is not solely
confined to \zfour\ but arises at lower redshift. To investigate this
briefly, we compared the \daoxrq\ and \daoxrl\ distributions of
$z=3$--4 HRLQs (9 objects) in the full sample of M11 with those of
$z<3$ HRLQs (274 objects), and found they are different at a
$\simgt5\sigma$ level (see Table~\ref{twostz_table}; $z=3$--4 HRLQs
have both larger \daoxrq\ and \daoxrl\ values
which indicate an \xray\ emission enhancement by a factor of
$\approx3$). However, the \daoxrq\ values of $z=2$--3 HRLQs (39
objects) follow a similar distribution to that of $z<2$ HRLQs (235
objects; $0.18\sigma$ level difference). For the \daoxrl\ parameter, although
its distribution for $z=2$--3 HRLQs is different from that of $z<2$
HRLQs at a $\approx2\sigma$ level, the mean \daoxrl\ value of
$z=2$--3 HRLQs is slightly lower than that of $z<2$ HRLQs
($-0.030\pm0.019$ vs. $0.005\pm0.008$). Therefore, the above tests
suggest that the \xray\ emission enhancement of HRLQs begins to arise
at $z\approx3$. This statistically independent result from $z=3$--4 also supports
the robustness of our findings of a HRLQ \xray\ emission enhancement
at \zfour. Furthermore, after combining the samples of \zfour\ HRLQs and $z=3$--4
HRLQs, we find that $z>3$ HRLQs have an \xray\ emission enhancement
over $z<3$ HRLQs at a $>5.7\sigma$ significance level corresponding to
a null-hypothesis probability of $\approx10^{-8}$ (also see Table~\ref{twostz_table}).

\subsubsection{Relevant Physical Considerations Regarding the X-ray Emission Enhancement}\label{discuss:daox:physical}

Our results suggest that the most-luminous relativistic jets generate
stronger \xray\ emission in the early universe. Previous studies by
Bassett et al.~(2004) and Lopez et al.~(2006) that included a mix of HRLQs
(including a subset of those in our sample) and moderately radio-loud
quasars found that high-redshift objects did not show substantially
stronger \xray\ emission compared to a lower-redshift sample of RLQs
detected by {\it ROSAT\/} (from Brinkmann et al.~1997). We assess further
these previous results by performing a similar two-sample analysis on the
moderately radio-loud quasars ($1<\log{R}<2.5$) in the full sample of M11
(which includes the moderately radio-loud objects from Bassett et al.~2004
and Lopez et al.~2006).  The \daoxrq\ distribution of \zfour\ moderately 
radio-loud quasars (10 objects; all \xray\ detected) is different from 
that of $z<4$ moderately radio-loud quasars (280 objects with 48 \xray\ 
upper limits) at a $\approx2\sigma$ level, while the \daoxrl\ values of 
\zfour\ moderately radio-loud quasars follow a similar distribution to 
that of $z<4$ moderately radio-loud quasars ($0.57\sigma$ level difference). 
Therefore, there is
no firm evidence that moderately radio-loud quasars have a significant
\xray\ emission enhancement compared to those at lower redshift
(recall also from \S\ref{intro} that RQQs show no evolution in \xray\
properties with redshift). This suggests that the \xray\ enhancement
in high-redshift HRLQs occurs within the jet-linked component, which likely
(e.g., M11) contributes an increasing fraction of the \xray\
continuum with increasing radio-loudness values.\footnote{It is
  obviously also
possible, but not required, that the jet-linked \xray\ generation
mechanism of HRLQs differs from (and is more sensitive to redshift than)
that of moderately radio-loud quasars.}  

The enhancements in relative nuclear \xray\ emission we
find for $z>4$ HRLQs are less than would be predicted for an
\xray\ jet-linked component dominated by IC/CMB emission, which
possesses a strong $(1+z)^{4}$ dependence on redshift (e.g., Schwartz
2002; see $\S$1). For example, the median redshift for the RLQs from
Miller et al.~(2011) with $z<4$, $\log R > 2.5$, and $m_{\rm i}<20$ is
$z=1.3$. If the jet-linked contribution to the nuclear \xray\
emission is at least $\approx$~50\% at $z=1.3$, as is a conservative
estimate for RLQs with $\log R > 2.5$ (e.g., Zamorani et al.~1981; Figure
7 of M11), then for a pure IC/CMB \xray\ jet the overall
enhancement in the nuclear \xray\ emission at $z=4.4$ would be
$\simgt$16, corresponding to an increase in \hbox{${\Delta}{\alpha}_{\rm ox,RLQ}$} 
of 0.46. However, our $z>4$ HRLQs only show an increase in
${\Delta}{\alpha}_{\rm ox,RLQ}$ of 0.15--0.20 (see Table~5). Instead,
our results are consistent with an \xray\ jet-linked component
in which the fractional contribution from the IC/CMB process at
\hbox{$z=1.3$} is $\simlt$6\% of the nuclear \xray\ emission, with the
balance of the jet-linked \xray\ emission arising from inverse
Compton scattering of other seed photon fields such as radiation from
the broad-line region or dust (e.g., Sikora et al.~2009), from the jet
synchrotron emission itself (e.g., Sokolov et al.~2004; Meyer et
al.~2012 find that synchrotron self-Compton emission is relatively
less important in high-powered jets), or from non-cospatial
synchrotron radiation in a structured jet (e.g., Migliori et
al.~2012). The overall enhancement to the nuclear \xray\
continuum then rises more gradually with redshift; at $z=4.4/3.0/2.0$
RLQs would be \xray\ brighter by factors of $\simlt$2.8/1.5/1.1
(relative to a $z=1.3$ template), corresponding to
${\Delta}{\alpha}_{\rm ox,RLQ}$ increases of
$0.17/0.07/0.02$.

Another possible explanation for the \xray\ 
  enhancement is that an increasing contribution of the \xray\
  emission from the jets is due to the inverse-Compton scattering of the
  photon field of the host galaxy; e.g., Hardcastle \& Croston (2011)
  provided a detailed modeling of upscattering of host-galaxy
  photon fields into high-energy emission. The \xray\ enhancement of
  our sample can be explained if the host galaxies of high-redshift 
  HRLQs are more luminous in the infrared/optical band than those at
  lower redshift. This higher infrared/optical luminosity could be
  generated by the enhanced star-formation activity of high-redshift
  host galaxies (e.g., Wang et~al. 2008, 2011a; Mor
  et~al. 2012). Smail et~al. (2012) found that for the high-redshift
  ($z=$3.6--3.8) radio galaxies in their study, the luminosity of the 
  extended \xray\ halo is correlated with the far-infrared luminosity
  of the galaxy, which supports the \xray\ generation mechanism of
  inverse Compton scattering of the far-infrared photons produced by
  star-formation activity. Since 
  the cosmic star formation rate density has a broad plateau starting
  at $z\sim1$ and running to $z\sim4$ (e.g., Hopkins \& Beacom 2006),
  one might expect in this model the enhancement to emerge at $z\sim1$ rather than
  at $z\sim$3--4. However, the generally massive host galaxies of
  quasars may show stronger evolution effects at earlier cosmic times
  owing to cosmic downsizing (e.g. Cowie et al. 1996; Panter et
  al. 2007). For example, Archibald et~al. (2001) found a $(1+z)^3$
  evolution of the total infrared luminosity of radio galaxies at high
  redshifts.

The \aox\ and \daox\ parameters are defined based on the \xray\ flux
density at rest-frame 2~keV, which corresponds to observed-frame
\hbox{$\sim0.3$--$0.4$~keV} for our \zfour\ HRLQs. However, the
rest-frame 2~keV flux density was derived from the observed-frame
0.5--2.0~keV count rate (see \S\ref{xray}), which means that we have
extrapolated the \xray\ spectrum to rest-frame 2~keV assuming a single
power-law model. It is known that some \zfour\ HRLQs show \xray\
spectral flattening below observed-frame 0.5~keV, i.e., an \xray\ flux
deficit compared to a single power-law model (e.g., Yuan
et~al. 2006). Therefore, the extrapolation process may have
over-estimated the \xray\ flux density at rest-frame 2~keV. Although
\xray\ spectral flattening has also been found in a few $z<3$ objects
(e.g., Page et~al. 2005), it would not affect the \xray\ flux
calculation because rest-frame 2~keV is covered by the observed-frame
0.5--2.0~keV band for these objects. To test whether the \xray\
emission enhancement we found for \zfour\ HRLQs is caused by this kind
of \xray\ spectral curvature, we recalculated the \aox\ and \daox\
values for our \zfour\ HRLQs utilized in the two-sample analyses above
(i.e., those with $m_i<20$) from the count rates in the observed-frame
ultrasoft band (\hbox{0.3--1.0~keV}; see results in
Table~\ref{us_table}) except for the two \swift\ sources
(PMN~J1155$-$3107 and PMN~J1951$+$0134) which do not have adequate
counts in the ultrasoft band due to the limited sensitivity of the
\swift\ XRT. These $\alpha_{\rm ox, US}$ values obtained from the
ultrasoft band are similar to those we have obtained in \S\ref{xray};
the mean difference between them is $\langle\alpha_{\rm ox,
  US}-\alpha_{\rm ox}\rangle=-0.03\pm0.01$. If we use the \daox\
values calculated from the ultrasoft band ($\Delta\alpha_{\rm ox, RQQ,
  US}$ and $\Delta\alpha_{\rm ox, RLQ, US}$) in the two-sample tests
above, the \daox\ distributions of \zfour\ HRLQs and $z<4$ HRLQs are
still different at a $\gtrsim3\sigma$ level. Therefore, \xray\
spectral curvature only contributes to a small fraction, if any, of
the \xray\ emission enhancement of \zfour\ HRLQs over those at lower
redshift. In fact, some of the \zfour\ HRLQs without the apparent
\xray\ spectral curvature (e.g., GB~1508$+$5714; Yuan et~al. 2006) are
among the objects we find to have the strongest \xray\ enhancements.  

Considering the seven \zfour\ HRLQs with high-resolution radio images (see \S\ref{sample}), the five with extended 
radio emission on mas scales have higher \daoxrq\ values \hbox{($\langle\Delta\alpha_{\rm ox, RQQ}\rangle=0.51\pm0.09$)} on average 
than those of the two without extended radio emission \hbox{($\langle\Delta\alpha_{\rm ox, RQQ}\rangle=0.15\pm0.24$)}. 
This might be because the objects with extended radio emission have higher average radio loudness 
(see \S\ref{sample}). However, After considering the strength of their
radio emission, the objects with extended radio emission still have 
a higher average \daoxrl\ value \hbox{($\langle\Delta\alpha_{\rm ox, RLQ}\rangle=0.23\pm0.09$)} than that of the objects 
without extended radio emission \hbox{($\langle\Delta\alpha_{\rm ox,
    RLQ}\rangle=-0.07\pm0.23$)} although with substantial uncertainty
due to the limited number of objects.  

\subsection{Optical/UV Broad Emission Lines}\label{discuss:lines} 

The optical/UV spectra of quasars at \zfour\ usually show prominent broad emission lines such as \lyanv\ 
and \civ\ (see Fig.~\ref{allspec_fig}). For RLQs, if the optical/UV continuum emission is also beamed following  
the radio emission, one might expect that the emission-line rest-frame equivalent widths (REWs) would be
correlated with the beaming of the radio emission. Therefore, we investigate the relation between emission-line REWs 
and radio loudness for our sample of HRLQs, and compare this relation to that for the majority of RLQs over a wide 
range of redshift. 

\begin{figure*}[t]
    \centering
    \includegraphics[width=5.5in]{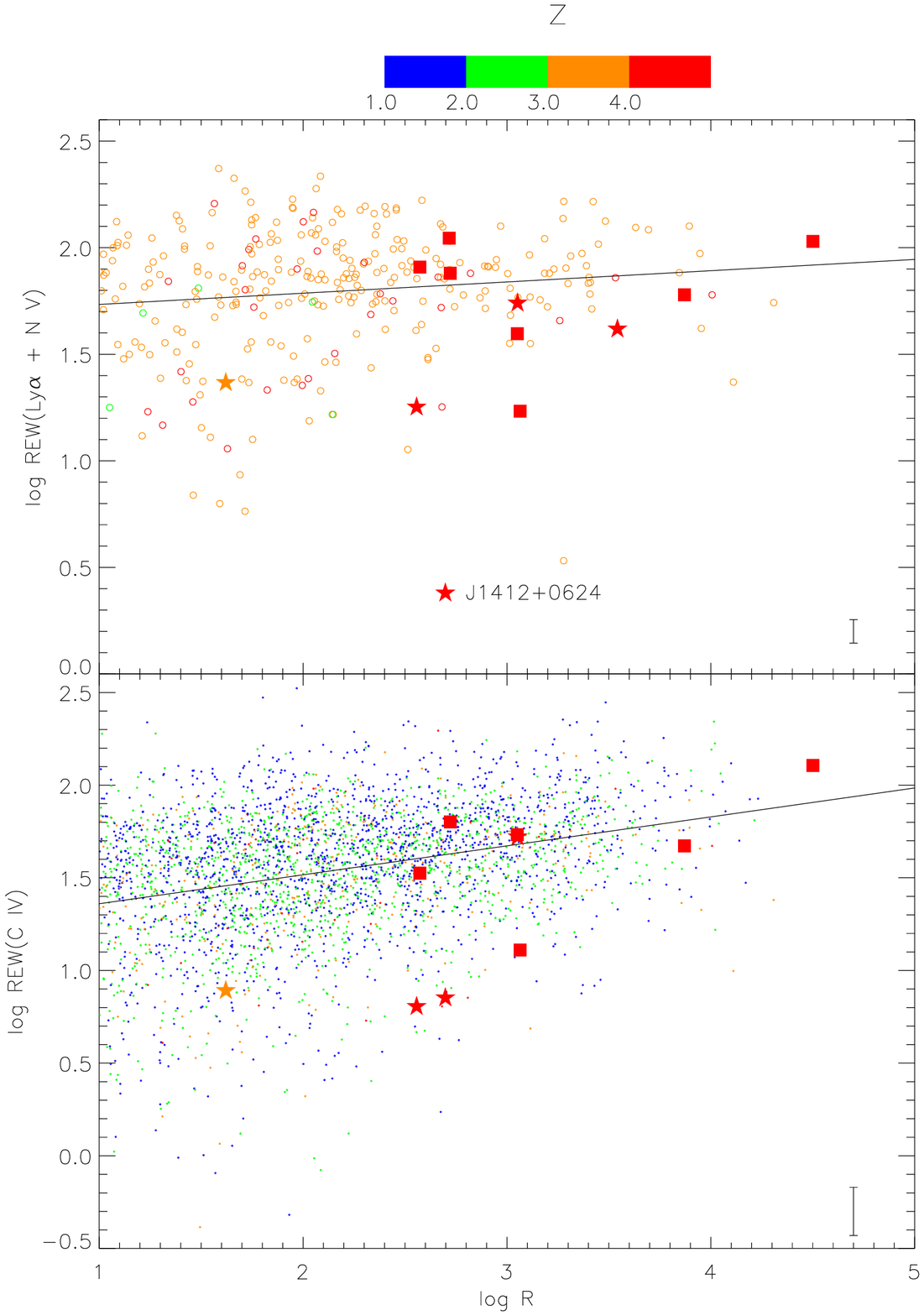}
    \caption{\footnotesize{The REWs of the broad emission lines \lyanv\ (upper panel) and \civ\ (lower panel) as a function of radio loudness. 
              In the upper panel, the open circles represent the radio-loud SDSS DR5 quasars in Table~1 of Diamond-Stanic et~al. (2009) which gives 
              the REW measurements of \lyanv. In the lower panel, the small dots represent the radio-loud SDSS DR7 quasars with \civ\ REW 
              measurements from Shen et~al. (2011). Our objects with available line REW measurements are shown in both panels as filled 
              stars (for \chandra\ Cycle~12 targets) or filled squares (archival objects). The weak-line quasar in our HRLQ sample (J1412$+$0624) is labeled 
              in the upper panel. The typical error bars for emission-line REWs of our objects are shown at the bottom-right corner of each panel. 
              All symbols are color-coded based on their redshifts using the color bar at the top of the figure.}
             \label{ewr_fig}}
\end{figure*}

For our \zfour\ HRLQs with SDSS spectroscopy, we obtained measurements of REWs from Diamond-Stanic et~al. (2009) 
and Shen et~al. (2011) for \lyanv\ and \civ, respectively (see Table~\ref{ew_table}).\footnote{We also list the values of
J1639$+$4340 in Table~\ref{ew_table} and the following Table~\ref{ntd_table} for reference despite it is a moderately radio-loud quasar.} 
Two objects (J1026$+$2542 and J1420$+$1205) with SDSS spectroscopy do not have REW(\lyanv) measurements in Diamond-Stanic et~al. 
(2009).\footnote{Diamond-Stanic et~al. (2009) measured the REW(\lyanv) values for SDSS DR5 quasars (Schneider et~al. 2007) 
at $z>3$. These two objects are in the SDSS DR7 quasar catalog but not in the DR5 quasar catalog.} We measured 
REW(\lyanv) for these two sources following the method in \S2 of Diamond-Stanic et~al. (2009). Available REW measurements 
for other objects without SDSS spectroscopy were also obtained from the literature (see Table~\ref{ew_table}).  
Measurements of REW(\lyanv) are usually affected by the presence of the Ly$\alpha$ forest, so our values likely underestimate 
the intrinsic strength of Ly$\alpha$. Although the number density of Ly$\alpha$ absorbers is known to evolve with redshift 
(e.g., Weymann et~al. 1998; Songaila \& Cowie 2002; Janknecht et~al. 2006), this effect will not introduce biases to our 
correlation analyses 
of emission-line REWs with radio loudness assuming the RLQ radio-loudness distribution does not evolve with redshift. 
The number of objects in our sample that have emission-line REW measurements is not adequate for reliable correlation analysis. 
Therefore, we first investigate these correlations for a sample of RLQs across a wide range 
of redshift, and then see whether our \zfour\ HRLQs follow similar trends. We selected all SDSS DR7 RLQs 
which have REW(\lyanv) and REW(\civ) measurements from Diamond-Stanic et~al. (2009) and Shen et~al. (2011), respectively. 
Fig.~\ref{ewr_fig} shows the relation between emission-line REWs and radio loudness, for which highly significant 
positive correlations are found (see Table~\ref{corr_table} for the correlation probabilities). Fig.~\ref{ewr_fig} also shows the 
best-fit linear correlation in the logarithmic parameter space using the IDL {\sc linfit} procedure, 
\begin{equation}
\log {\rm REW(Ly\alpha+N\ V)}=(1.68\pm0.06)+(0.05\pm0.02)\log R,
\end{equation}
and
\begin{equation}
\log {\rm REW(C\ IV)}=(1.20\pm0.02)+(0.16\pm0.01)\log R.
\end{equation}

Our HRLQs generally follow the above positive correlations between emission-line REWs and radio-loudness (see the filled stars and 
squares in Fig.~\ref{ewr_fig}), showing that these correlations do not seem to evolve with redshift. These positive correlations are 
consistent with the results in Kimball et~al. (2011a) on REW(Mg~{\sc ii}) and REW(\civ) for SDSS~DR5 quasars.\footnote{Kimball et~al. 
(2011a) found positive correlations between the core radio-to-optical flux ratio $R_{\rm I}$ and the REW values of Mg~{\sc ii} and \civ\ 
for SDSS RLQs. The definition of $R_{\rm I}$ is generally different from that of radio loudness because the latter is 
defined based on the total radio luminosity. However, the positive correlations in Kimball et~al. (2011a) also hold true for their 
subsample of RLQs without extended radio emission (the ``core'' morphology class which makes up $\approx70\%$ of their 
full sample; see their Table~8). In this case, the definition of $R_{\rm I}$ is similar to that of radio loudness. Therefore, our results 
are consistent with those of Kimball et~al. (2011a).} 
One possible explanation for them is that the broad emission-line  
region of RLQs may emit anisotropically (see \S6.2 of Kimball
et~al. 2011a). Additionally or alternatively, the decrease in
synchrotron peak frequency with increasing peak or kinetic luminosity
in FSRQs (e.g., Meyer et al.~2011, their Figure 4) might more
stringently limit jet dilution of optical emission-line features in
more radio-loud quasars. 
Our results indicate that for HRLQs with broad 
emission lines (i.e., not BL Lac objects), the beamed relativistic jets generally have negligible contribution to the optical/UV continua 
(except for a few outliers with exceptionally weak optical/UV emission lines; see below). 
This is consistent with recent results from SED studies of HRLQs (e.g., Ghisellini et~al. 2010) that their optical/UV emission 
is disk dominated. Furthermore, this finding of a minimal jet-linked contribution to 
the optical/UV continuum validates the use of \daoxrq\ as a diagnostic quantity for our HRLQs; note in Column (16) of Table~\ref{aox_table} 
that the \aox--$L_{2500\mbox{\rm~\scriptsize\AA}}$ correlation for RQQs is utilized.

To assess further the jet contribution to the optical/UV continua of our HRLQs, we calculated the non-thermal dominance parameter 
(NTD $\equiv L_{\rm obs}/L_{\rm pred}$) of the optical/UV continuum following 
Shaw et~al. (2012). $L_{\rm obs}$ is the observed monochromatic continuum luminosity near an emission line ($\nu L_{1350}$ near \civ\ in our 
case), while $L_{\rm pred}$ is the predicted $\nu L_{1350}$ from the relation between continuum luminosity and emission-line luminosity in 
Shen et~al. (2011; see their Equation 14) obtained from typical quasars. $L_{\rm pred}$ represents the optical/UV continuum 
emission from the accretion disk. NTD$\gg1$ means that there is substantial jet-contributed optical/UV continuum. We calculated the 
NTD values for our HRLQs with SDSS spectroscopy covering the \civ\ line (see Table~\ref{ntd_table}). Three objects (J1420$+$1205, 
J1235$-$0003, and J1325$+$1123) have NTD values close to unity, supporting our results that the relativistic jets of our HRLQs generally have little 
contribution to their optical/UV continua. Another object, GB~1508$+$5714 (${\rm NTD}=1.3$), may possibly have some jet-linked optical/UV 
continuum emission. However, its NTD value is consistent with unity considering the scatter of the relation between $\nu L_{1350}$ and 
$L_{\rm C~IV}$ ($\sim0.2$ dex; see \S3.7 of Shen et~al. 2011). The other two HRLQs (J1412$+$0624 and J1659$+$2101) have ${\rm NTDs}\approx6$ 
due to their exceptionally weak \civ\ lines that are discussed further below. 

For the REW(\lyanv)--$\log R$ correlation, it appears that there are a few outliers with exceptionally weak \lyanv\ lines (see the upper 
panel of Fig.~\ref{ewr_fig}); they were identified as weak-line quasars (WLQs) by Diamond-Stanic et~al. (2009), including one of our \chandra\ 
Cycle~12 targets (J1412$+$0624). Another \zfour\ HRLQ (J1659$+$2101) in our sample has REW(\lyanv) values (17.9~\AA; see Table~\ref{ew_table}) 
close to the WLQ criterion of \hbox{REW(\lyanv)$ < 15.4$~\AA}. These two objects also have 
weak \civ\ emission lines. Their REW(\civ) values are below the $3\sigma$ negative deviation from the mean REW(\civ) of SDSS quasars 
(Wu et~al. 2012). The weak emission lines of these objects may suggest they have a relativistically 
boosted optical/UV continuum which dilutes the emission lines. However, the broad-band SEDs of these objects (see \S\ref{discuss:sed}) 
do not show the signature of a relativistically boosted continuum (e.g., a parabolic profile of the SED; Nieppola et~al. 2006). It is also possible that the 
weakness of their emission lines is caused by extreme quasar disk-wind properties as for the radio-quiet weak-line quasars (e.g., Richards et~al. 2011; 
Wu et~al. 2011, 2012). One of our moderately radio-loud quasars (J1639$+$4340) also has a 
weak \civ\ line (see Table~\ref{ew_table}). This object is more likely to fit in the latter scenario; and thus the apparent high NTD value of 
this object does not indicate that it has substantial jet-contributed optical/UV continuum. 

\begin{figure*}[t]
    \centering
    \includegraphics[width=5.5in]{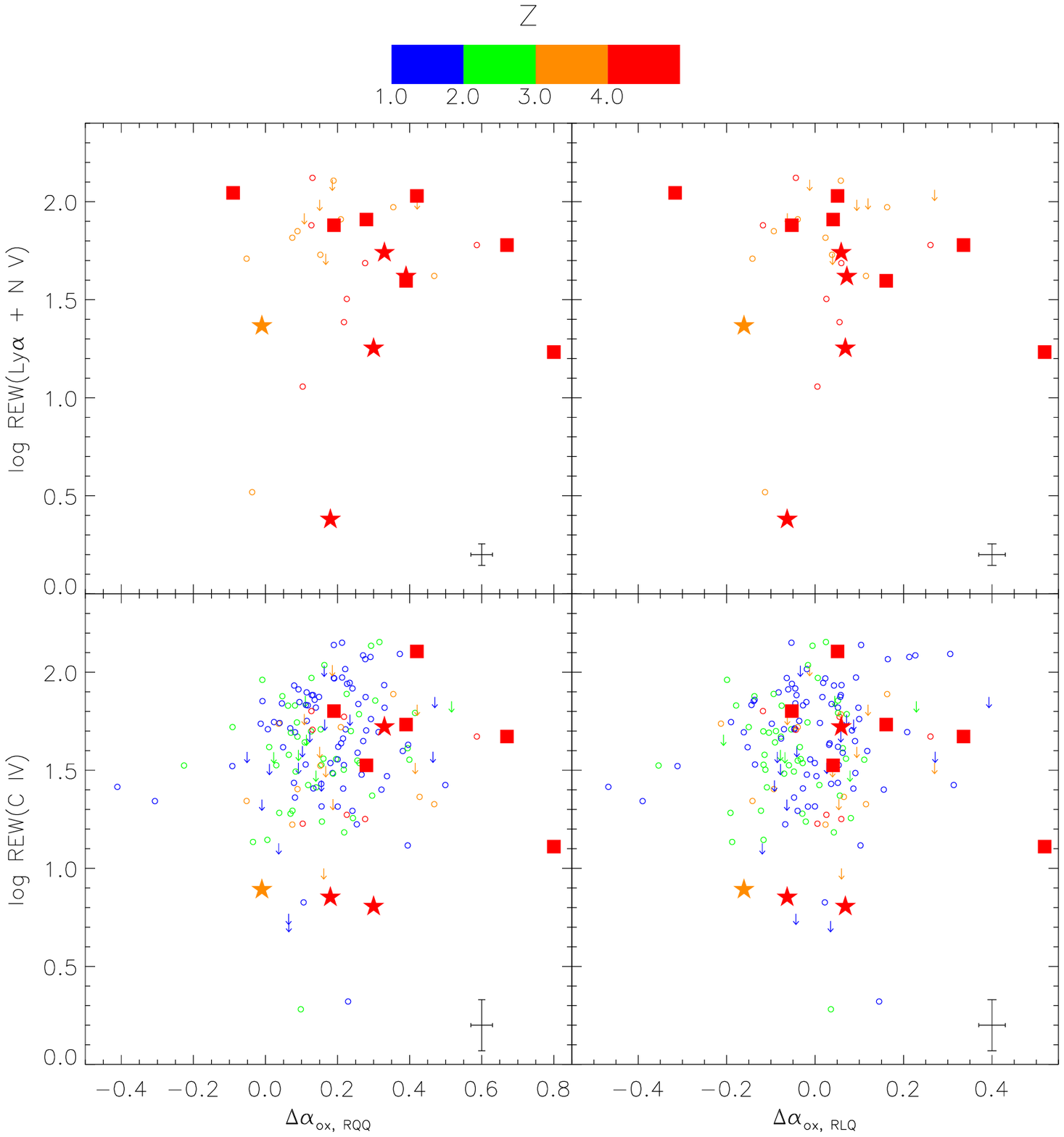}
    \caption{\footnotesize{The REWs of the broad emission lines \lyanv\ (top row) and \civ\ (bottom row) as a function of \daoxrq\ 
              (left column) and \daoxrl\ (right column), respectively. The open circles in the top-row panels represent the 
              M11 objects with available REW(\lyanv) measurements in Table~1 of Diamond-Stanic et~al. (2009). In the 
              bottom-row panels, the open circles are the M11 objects with REW(\civ) 
              measurements from Shen et~al. (2011). Our objects with available line REW measurements are shown in both panels as filled 
              stars (for \chandra\ Cycle~12 targets) or filled squares (archival objects). The typical error bars for emission-line REWs, \daoxrq, and \daoxrl\ 
              of our objects are shown at the bottom-right corner of each panel. All symbols are color-coded based on their redshifts 
              using the color bar at the top of the figure.}
             \label{ewdaox_fig}}
\end{figure*}

The jet-linked \xray\ emission of RLQs 
is also likely beamed, although probably to a lesser extent compared to jet radio emission (e.g., see \S6 of M11). 
Therefore, the relation between emission-line REWs and relative \xray\ brightness is also worth investigating.
A correlation analysis was performed on a sample combining our RLQs and the objects in M11 that 
have available \hbox{\lyanv} and \civ\ REW measurements (see Fig.~\ref{ewdaox_fig} for the relation between line REWs and relative 
\xray\ brightness of this sample) using Spearman's rank-order analysis in the ASURV package; Spearman's rank-order analysis is usually 
preferred over Kendall's $\tau$ test for samples where the number of objects is $N\geqslant30$.\footnote{See Appendix A3.3 of the 
ASURV manual at http://astrostatistics.psu.edu/statcodes/asurv. In our work, we choose Spearman's rank-order analysis when sample 
size $N\geqslant30$ and Kendall's $\tau$ test when $N<30$.} 
The results are shown in Table~\ref{corr_table}. Only a marginal correlation ($93.3\%$) was found between REW(\civ) and \daoxrq, 
which is consistent with the result in Richards et~al. (2011) that RLQs with weaker \civ\ lines also tend to be 
weaker in \hbox{X-rays}. However, the bottom-left panel (\daoxrq\ vs. \civ\ REW) of Fig.~\ref{ewdaox_fig} 
shows a large scatter in this correlation. No other significant correlations were found between relative \xray\ brightness 
and emission-line strength for the RLQs (see Table~\ref{corr_table} and Fig.~\ref{ewdaox_fig}). 

\subsection{X-ray Spectral Properties}\label{discuss:spec}

\begin{figure*}[t]
    \centering
    \includegraphics[width=5.5in]{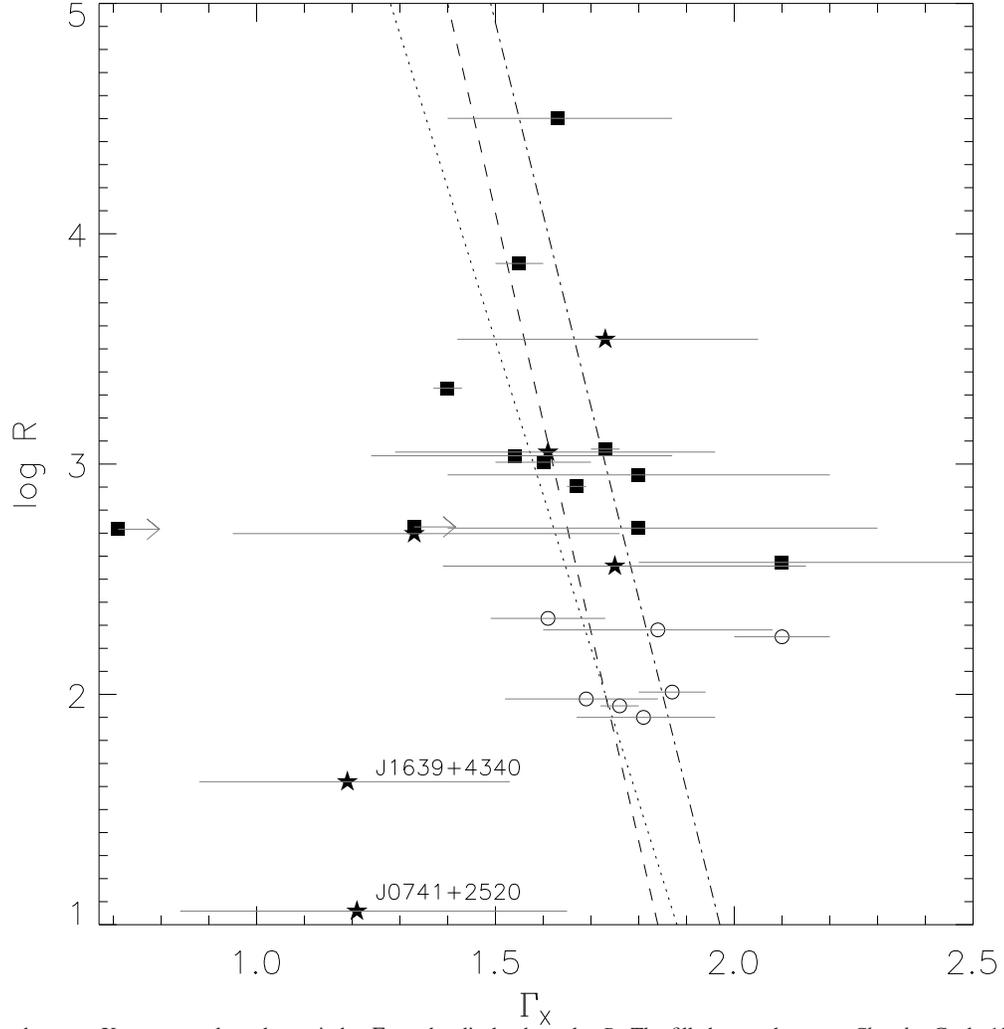}
    \caption{\footnotesize{The relation between X-ray power-law photon index $\Gamma_{\rm X}$ and radio loudness $\log R$. The filled
              stars show our \chandra\  Cycle 12 targets, while the filled squares show our HRLQs with 
              sensitive archival \xray\ coverage. The open circles represent \zfour\ moderately radio-loud 
              quasars in Saez et~al. (2011). The grey horizontal lines (rightward arrow) represent the $1\sigma$ 
              error bars (lower limit) for the \xray\ power-law photon indices. The dashed line shows our best-fit 
              correlation between $\Gamma_{\rm X}$ and $\log R$ for $z\gtrsim4$ RLQs, while the dotted line and dash-dotted line represent the 
              \hbox{$\Gamma_{\rm X}$--$\log R$} correlations for $z>2$ RLQs in Saez et~al. (2011) and for $z<2$ RLQs in Reeves \& Turner (2000), 
              respectively. The two moderately radio-loud objects among our \chandra\ Cycle~12 targets are apparent outliers (J0741$+$2520 and 
              J1639$+$4340; see the two filled stars labeled in the bottom left part of the figure).}
             \label{gamma_fig}}
\end{figure*}

Our \chandra\ Cycle~12 targets do not have sufficient counts for individual \xray\ spectral analysis, 
and thus we obtained their effective power-law photon indices based on their band ratios (see \S\ref{xray}). 
The \xray\ power-law photon index for RLQs is known to be anti-correlated with $\log R$ (e.g., Wilkes 
\& Elvis 1987; Reeves \& Turner 2000; Lopez et~al. 2006; Saez et~al. 2011). 
Our sample of HRLQs at $z>4$ is generally consistent with this anti-correlation (see Fig.~\ref{gamma_fig}). 
Kendall's $\tau$ test on a sample combining our HRLQs and the moderately radio-loud quasars at \zfour\ in Saez et~al. (2011) shows a $2\sigma$-level 
anti-correlation between $\Gamma_{\rm X}$ and $\log R$ (95.6\% correlation probability). The best-fit correlation for this sample, obtained 
via the EM linear-regression algorithm in the ASURV package,\footnote{This linear-regression algorithm is applicable to censored data. Our 
sample has two objects with upper limits upon $\Gamma_{\rm X}$.} is
\begin{equation}
\Gamma_{\rm X} = (1.95\pm0.28)+(-0.11\pm0.10)\log R.
\end{equation}
This $\Gamma_{\rm X}$--$\log R$ relation (the dashed line in Fig.~\ref{gamma_fig}) is consistent with that for the $z>2$ RLQs in Saez et~al. 
(2011; see their Equation 1) and with that for the $z<2$ RLQs in Reeves \& Turner (2000; see Equation 2 of Saez et~al. 2011), suggesting that 
this relation does not evolve with redshift. The two moderately
radio-loud quasars observed in \chandra\ Cycle~12 (J0741$+$2520 and
J1639$+$4340), which do not satisfy our criterion for HRLQs, 
are outliers from this correlation. Instead of having the softer \xray\ spectra generally associated 
with lower radio loudness, they have the hardest spectra among our objects. 

HRLQs often appear to have intrinsic \xray\ absorption with $N_H\gtrsim10^{22}$~cm$^{-2}$ (e.g., Yuan 
et~al. 2006). Bassett et~al. (2004) found evidence of a similar level of intrinsic \xray\ absorption 
for \zfour\ moderately radio-loud quasars. To assess possible indications of intrinsic absorption for our 
\chandra\ Cycle 12 targets, we performed joint spectral analyses. We divided our six targets into two 
groups with $\log R > 2.5$ (four objects) and $\log R < 2.5$ (2 objects), respectively. For each object, the \xray\ 
spectrum in observed-frame \hbox{0.5--8.0~keV} (corresponding to rest-frame \hbox{$\approx2$--40~keV}) was extracted from a 
$3''$-radius circular region centered on the \xray\ position using the standard CIAO routine 
{\sc psextract}. Background spectra were extracted from annular regions with inner and outer radii of $6''$ and $9''$, 
respectively. All background regions are free of \xray\ sources. Joint \xray\ spectral fitting was performed with XSPEC 
v12.6.0 (Arnaud 1996). We applied the $C$-statistic (Cash 1979) in the spectral fitting instead of
the standard $\chi^2$ statistic because the $C$-statistic is well suited to limited \xray\
count scenarios (e.g., Nousek \& Shue 1989). We fit the spectra jointly using two models: (1) a power-law
model with a Galactic absorption component represented by the \verb+wabs+ model (Morrison
\& McCammon 1983); (2) another model similar to the first, but adding an intrinsic 
(redshifted to the source rest frame) neutral absorption component, represented by the \verb+zwabs+ model. Each source
was assigned its own values of redshift and Galactic neutral hydrogen column density (see Column~4 of Table~\ref{aox_table}). 
Table~\ref{xspec_table} shows the best-fit spectral parameters along with the errors or upper limits at the 90\%
confidence level for one parameter of interest ($\Delta\;C=2.71$; Avni 1976; Cash 1979).
The best-fit power-law photon index values are consistent with those from band-ratio analysis (see Table~\ref{cts_table}). 
Our joint-fitting procedure is not able to provide tight constraints on the intrinsic \xray\ absorption for either highly or 
moderately radio-loud quasars (see the upper limits on intrinsic $N_H$ in Table~\ref{xspec_table}) mainly due to the limited 
number of \xray\ counts and their high redshifts. Adding an intrinsic absorption component does not improve the quality of the joint fits. 

As discussed above, our two moderately radio-loud quasars appear to have a harder average \xray\ spectrum than that of 
our HRLQs, which appears inconsistent with the known anti-correlation between \xray\ power-law photon index and 
radio loudness. Possible heavy \xray\ absorption may exist in these two objects. We estimate the potential intrinsic 
\xray\ absorption for these two moderately radio-loud quasars by assuming their unabsorbed power-law photon indices follow the 
\hbox{$\log R$--$\Gamma_{\rm X}$} relation (see Equation 8; $\Gamma_{\rm X, unabs}=1.79$ for J0741$+$2520, $\Gamma_{\rm X, unabs}=1.74$ for J1639$+$4340), 
and then adding an intrinsic \xray\ absorption component to make the apparent $\Gamma_{\rm X}$ agree with the values obtained from the band-ratio analysis
($\Gamma_{\rm X}=1.21$ for J0741$+$2520, $\Gamma_{\rm X}=1.19$ for J1639$+$4340; see Table~\ref{cts_table}). The required intrinsic hydrogen column 
densities are \hbox{$N_{\rm H}=3.3\times10^{23}$~cm$^{-2}$} for J0741$+$2520 and \hbox{$N_{\rm H}=2.0\times10^{23}$~cm$^{-2}$} for J1639$+$4340, which are 
about an order of magnitude higher than the values found for high-redshift RLQs in previous studies. The intrinsic \xray\ absorption in RLQs is often associated with 
absorption features in their optical/UV spectra (e.g., Elvis et~al. 1998). The optical/UV spectrum of J1639$+$4340 (see the top-right panel of 
Fig.~\ref{allspec_fig}) shows several \civ\ and Si~{\sc iv} absorption features, although none of them has sufficient width to be identified as 
a broad absorption line (BAL). The optical/UV spectrum of J0741$+$2520 (see Fig.~3 of McGreer et~al. 2009) does not show Si~{\sc iv} absorption 
features; it does not have coverage of the \civ\ region. 

\subsection{The Radio-to-$\gamma$-ray Spectral Energy Distributions}\label{discuss:sed}

To investigate the broad-band SEDs of our RLQs, we gathered photometric data from the following sources:
\begin{enumerate} 
\item Radio: The 1.4~GHz flux densities are from the FIRST or NVSS surveys; the 5~GHz values are from the GB6 or PMN 
catalogs or individual observations (see \S\ref{xray}); the flux densities at other frequencies were retrieved from the NED.
\item Mid-infrared: From the all-sky catalog of the {\it Wide-field Infrared Survey Explorer} ({\it WISE}; Wright et~al. 
2010).\footnote{See the {\it WISE} all-sky data release at http://wise2.ipac.caltech.edu/docs/release/allsky/.} All 
of our objects have {\it WISE} detections in at least two bands (W1 at $3.4\ \mu$m and W2 at $4.6\ \mu$m) except for 
GB~1428$+$4217 and PMN~J1951$+$0134. We examined the {\it WISE} images of these two objects. There appears to be a source at the optical 
location of GB~1428$+$4217, but it is strongly blended with a nearby brighter source $9''$ away. There is no source at the optical location of 
PMN~J1951$+$0134 in the relevant {\it WISE} images. All other objects are free of issues such as source blending, confusion, or mismatching. 
The high detection fraction ($>17/19$) in the mid-infrared band is remarkable given the high redshifts of 
our objects.
\item Near-infrared: From the Two Micron All Sky Survey (2MASS; Skrutskie et~al. 2006). For the objects having SDSS spectroscopy, 
we also checked the SDSS DR7 quasar catalog which provides additional deeper 2MASS photometry. Only our two moderately radio-loud quasars 
(J0741$+$2520 and J1639$+$4340) have 2MASS detections, likely owing to their exceptional overall luminosities. GB~1428$+$4217 has additional 
near-infrared observations with the United Kingdom Infra-Red Telescope (UKIRT) reported by Fabian et~al. (1999). 
\item Optical: From the SDSS database and/or the NED. The photometry for the bands that 
are seriously affected by the Ly$\alpha$ forest is omitted. We do not include any UV photometric information for the same reason.  
\item X-ray: From this work. 
\item $\gamma$-ray: The upper limits upon $\gamma$-ray luminosities are derived from the two-year survey data of the Large Area Telescope 
(LAT; Atwood et~al. 2009) onboard the {\it Fermi \hbox{Gamma-Ray} Space Telescope}. See below for details. 
\end{enumerate}
The radio-to-$\gamma$-ray SEDs of our objects are plotted in
Fig.~\ref{seds_fig}, ordered by $\log R$ in descending order. We
compare the SEDs of our high-redshift HRLQs to those of similar RLQs
at lower 
redshifts. There are two commonly used composite SEDs of RLQs in the literature:
Elvis et~al. (1994; E94 hereafter) and Shang et~al. (2011; S11
hereafter). The E94 SED is biased toward \xray\ bright quasars as they
required their objects to have \xray\ observations with good
signal-to-noise ratio (see discussion in \S3 of E94 and \S5.1 of
Richards et~al. 2006a). For the S11
SED, although their sample is not \xray\ selected, the majority of
their RLQs have lower optical/UV luminosities than those of our
objects. The optical/UV luminosity can affect the shape of the
broad-band SED (see Eqn. 5 for the dependence of \aox\ on optical/UV
luminosity). The \aox\ parameter of RLQs also depends on radio
luminosity (see Eqn. 5). Therefore we selected 10 RLQs in the S11 sample with
comparable optical/UV luminosity ($\log\lambda
L_{\lambda}$[3000~\AA]$>45.9$) and radio-loudness values ($2.9 < \log
R < 3.7$) to those of our \zfour\ HRLQs. We construct a comparison
composite SED using these 10 objects (see the solid lines in
Fig.~\ref{seds_fig}). We normalized the comparison SED to the observed data at rest-frame 2500~\AA\ for the following 
reasons: (1) rest-frame 2500~\AA\ is within a region without strong emission lines; (2) the optical/UV continuum generally does not 
have a substantial contribution from the jet (except perhaps for a few objects with weak emission lines; see \S\ref{discuss:lines}); 
and (3) when SEDs are normalized at rest-frame $2500$~\AA, the
relative positions of \xray\ data points directly reflect the \aox\
values of the relevant quasars. One caveat about the SEDs is that our multi-band
photometry is non-simultaneous, and some HRLQs are known to have strong
variability. 

The 10 RLQs from S11 used in constructing our comparison SED satisfy our $\log R$ criterion for HRLQs; they are HRLQs at low 
redshifts ($z<1.4$). The majority of our HRLQs (11/17) have higher \xray\ luminosities than those of the comparison SED. 
Meanwhile, they do not show weaker optical/UV emission relative to their infrared emission. Therefore, the generally higher 
\daox\ values of our \zfour\ HRLQs (see \S\ref{discuss:daox}) do reflect stronger \xray\ emission instead of fainter optical/UV emission. This 
is an additional illustration of the \xray\ emission enhancement of \zfour\ HRLQs compared to HRLQs at lower redshift. The SEDs 
of RX~J1028$-$0844, GB~1428$+$4217, and GB~1508$+$5714 all 
reach maximal $\nu L\nu$ values (over the $10^9$--$10^{19}$~Hz range) in the \xray\ band. This type of SED has been modeled 
as the emission generated by Compton scattering of synchrotron emission from the jet itself and/or of a powerful 
external photon field such as the thermal emission of the accretion process (SSC/EC; see Fabian et~al. 1998, 1999;
Ghisellini et~al. 1998). The SED of Q~0906$+$6930 has also been described with a similar model (Romani 2006). Q~0906$+$6930, 
along with PMN~J0525$-$3343 and GB~1713$+$2148, have notably stronger
mid-infrared emission than that of the comparison SED, especially in  
the longer wavelength bands (W3 at 12~$\mu$m and W4 at
22~$\mu$m). This mid-infrared excess may have a contribution from the  
jet synchrotron radiation (e.g., Ghisellini et~al. 1998). An
alternative origin for the mid-infrared excess is the thermal
emission of circumnuclear dust predicted by the clumpy torus model
(e.g., Nenkova et~al. 2008). The SEDs of our HRLQs are significantly different from those
of typical BL~Lac objects which generally have a parabolic profile peaking in the near-infrared band (e.g., Nieppola et~al. 2006). 

Many blazars are high-energy \garay\ emitters above 100~MeV (e.g., Hartman et~al. 1999). Q~0906$+$6930 has a tentative claimed detection 
($\sim1.5\sigma$) in its \garay\ observation by the Energetic Gamma-Ray Experiment Telescope (EGRET) onboard the {\it Compton Gamma-Ray 
Observatory} (Romani et~al. 2004). The {\it Fermi} LAT 
has an order of magnitude better sensitivity and positional accuracy than EGRET. Romani (2006) suggested Q~0906$+$6930 should be detected by 
the {\it Fermi} LAT in its first year of operation. However, although the first two-year survey of the {\it Fermi} LAT has detected 
$>800$ blazars at $z\lesssim3$ (e.g., Ackermann et~al. 2011), none of our HRLQs, including Q~0906$+$6930, is detected at present. 
We searched the second source catalog of the {\it Fermi} LAT (2FGL; Nolan et~al. 2012) and found no counterparts for our HRLQs within a $10^\prime$ 
matching radius.\footnote{We also searched for the \zfour\ HRLQs without sensitive \xray\ coverage listed in Table~\ref{noxray_table}, and found none of them 
having counterparts in the 2FGL catalog.} To obtain upper limits upon \garay\ luminosity, we first estimated the photon flux limits (in units of 
photons~cm$^{-2}$~s$^{-1}$) for all objects based on the point-source flux limit map of the {\it Fermi} LAT two-year survey 
(see Fig.~1 of Ackermann et~al. 2011).
We converted the photon flux limits to energy flux limits between observed-frame 
100~MeV--100~GeV and then flux-density limits at rest-frame 1~GeV, assuming a typical power-law photon index of {\it Fermi}-detected 
FSRQs ($\Gamma_\gamma=2.4$; see Fig.~18 of Ackermann et~al. 2011). The upper limits upon the monochromatic \garay\ luminosity at rest-frame 
1~GeV, calculated from the flux-density limits at the same frequency, are in the range of (1.2--2.6)$\times10^{47}$~erg~s$^{-1}$~Hz$^{-1}$ 
(see Table~\ref{fermi_table} and Fig.~\ref{seds_fig}). 
For some of our \zfour\ HRLQs with strong \xray\ enhancements (e.g., Q~0906$+$6930, RX~J1028$-$0844, GB~1428$+$4217, and GB~1508$+$5714), the \garay\ 
luminosity limits appear inconsistent with the prediction of the synchrotron-Compton blazar emission model in Romani et~al. (2006; see their Fig.~3). 
Further {\it Fermi} LAT survey observations will provide tighter \garay\ luminosity upper limits 
(by a factor of $\sim2$) or perhaps detections of \zfour\ blazars, which will help constrain the emission models for blazars at high redshift. 
Sbarrato et~al. (2012a) found a positive correlation 
between the luminosity of optical/UV broad emission lines and the \garay\ luminosity for {\it Fermi}-detected blazars. This correlation 
may suggest that the \zfour\ blazars with higher optical/UV emission-line luminosity should be easier to detect with {\it Fermi}. 

Six objects with SDSS spectroscopy have available black-hole mass ($M_{\rm BH}$) and Eddington-ratio ($L/L_{\rm Edd}$) estimates (see Table~\ref{mbh_table}) 
in the catalog of Shen et~al. (2011).\footnote{Another HRLQ in our
  sample, J1026$+$2542, has $M_{\rm BH}$ estimated by Sbarrato
  et~al. (2012b) based on the peak frequency of disk emission and the
  total disk luminosity. The estimated range is $M_{\rm
    BH}$=(1.8--4.5)$\times 10^9M_\odot$. Although this object has SDSS
spectroscopy, its $M_{\rm BH}$ cannot be measured based on available
emission-line widths because none of the standard lines (H$\beta$, \mgii, and
\civ) is covered due to its high redshift ($z=5.304$).} The bolometric luminosity in Shen et~al. (2011) was calculated from the optical/UV luminosity and the bolometric 
correction from the composite SED of Richards et~al. (2006a). Therefore, the bolometric luminosity and the Eddington ratio only account for 
the emission of the quasar from its accretion activity (i.e., excluding beamed jet emission). All of the six objects have $M_{\rm BH} > 10^9M_\odot$ except for 
GB~1508$+$5714. These five objects 
have estimated Eddington ratios between $0.1$ and $1$, which are typical for SDSS quasars at high redshift (e.g., Shen et~al. 2008). GB~1508$+$5714 has a smaller 
$M_{\rm BH}$ but an Eddington ratio $L/L_{\rm Edd} \approx 3$. This object is similar to the {\it Fermi}-detected FSRQs in Shaw et~al. (2012) which have smaller 
$M_{\rm BH}$ and higher Eddington 
ratios than those of optically selected quasars. The objects in Shaw et~al. (2012) show significant non-thermal (synchrotron) emission in the optical band 
associated with the jet. However, GB~1508$+$5714 has a smaller NTD value (consistent with unity; see \S\ref{discuss:lines}) than the majority of the objects in Shaw 
et~al. (2012; see their Fig. 3), which suggests that this object does not have substantial jet-linked optical/UV continuum emission. 
It is worth noting that the $M_{\rm BH}$ measurements for our objects are all based on the \civ\ emission line due to their high redshifts. 
The \civ\ estimator for $M_{\rm BH}$ is more affected by the quasar disk wind than the H$\beta$ and Mg~{\sc ii} estimators, 
and this may lead to errors for individual $M_{\rm BH}$ measurements (e.g., Shen et~al. 2008). Thus, the apparent super-Eddington emission of 
GB~1508$+$5714 should not be over-interpreted.


\section{Summary and Future Studies}\label{summary}

We have compiled a sample of 17 highly radio-loud quasars ($\log R>2.5$) at \zfour\ with sensitive \xray\ coverage
by \chandra, \xmm, or \swift. Four of them were targeted in new \chandra\ Cycle~12 observations, while the other 13 
objects have sensitive archival \xray\ coverage. Our sample of HRLQs represents the top $\sim5\%$ of the total RLQ 
population in terms of radio loudness. They are among the most-luminous quasars in both the radio and 
optical/UV bands. The \xray\ and broad-band properties of our HRLQs are presented and investigated. We have also reported new 
\chandra\ Cycle~12 observations of two moderately radio-loud quasars ($1 < \log R < 2.5$) at $z\gtrsim4$ which are among the most 
optically luminous RLQs. Our main results are summarized as follows. 
\begin{enumerate}
\item All of our \chandra\ Cycle~12 targets are detected in
  \hbox{X-rays}. None of them shows detectable extended \xray\ 
emission. All of the archival objects 
in our sample of \zfour\ HRLQs were also detected by \chandra, \xmm, or \swift, including four objects for which 
their archival \chandra\ (GB~1713$+$2148), \xmm\ (SDSS~J1235$-$0003), or \swift\ (PMN~J1155$-$3107 and PMN~J1951$+$0134) 
observations are first reported in this work. See \S\ref{xray}.
\item Our HRLQs at \zfour\ show stronger \xray\ emission (by a typical factor of $\approx3$) than HRLQs at 
lower redshift with similar optical/UV and radio luminosities. 
This contrasts with the behavior of moderately radio-loud quasars at \zfour, implying that the
high-redshift \xray\ enhancement occurs within the jet-linked component that becomes increasingly prominent for HRLQs.
We examined possible biases in our analysis 
and found that our result remains robust after controlling for a variety of potential issues. 
A similar \xray\ emission enhancement is also found for $z=3$--4 HRLQs,
which provides statistically independent evidence for our
findings. The observed \xray\ enhancement is not likely to be caused by the \xray\ 
spectral curvature found in some high-redshift HRLQs. See \S\ref{discuss:daox}.
\item Our \zfour\ HRLQs are generally consistent with the positive correlations between optical/UV emission-line 
REWs (\lyanv\ and \civ) and radio loudness found for the typical RLQ population. These positive correlations suggest that 
the optical/UV continua of HRLQs usually have a negligible
contribution from the relativistic jets (except for perhaps 
a few outliers with weak emission lines). Our 
\zfour\ HRLQs do not show strong correlations between the relative \xray\ brightness and 
optical/UV emission-line REWs. See \S\ref{discuss:lines}.
\item Our sample of \zfour\ HRLQs generally follows the known anti-correlation between \xray\ power-law photon 
index and radio loudness. However, the two targeted moderately radio-loud quasars, having 
the hardest \xray\ spectra among our objects, appear to be outliers
from this correlation, suggesting the possible presence of intrinsic
\xray\ absorption \hbox{($N_{\rm H} \sim 10^{23}$~cm$^{-2}$)}.  
See \S\ref{discuss:spec}.
\item The majority of \zfour\ HRLQs have higher \xray\ luminosities
  than those of the matched low-redshift HRLQs used to construct a comparison SED, 
which further illustrates and supports the \xray\ emission enhancement of \zfour\ HRLQs over those at lower redshifts. The SEDs 
of our objects with the highest \xray\ luminosities are consistent with the expectations from SSC/EC models. Some of our 
HRLQs also show an excess of mid-infrared emission which may be contributed by synchrotron radiation from the 
relativistic jets. None of our HRLQs has been detected by the {\it Fermi} LAT in its two-year survey. See \S\ref{discuss:sed}.
\end{enumerate}

We have performed the first systematic \xray\ survey of HRLQs at $z>4$. Our primary
new result is the measurement of an increase (by a factor of
$\approx$~3) in the relative \xray\ 
emission of $z>4$ HRLQs compared to their low-redshift counterparts. This 
work adds the relative \xray\ brightness of HRLQs to the RLQ properties that evolve 
with redshift, such as the frequency of intrinsic \xray\ absorption (see $\S$1) 
and perhaps the RLQ fraction (e.g., Jiang et al.~2007; Singal et al.~2012). 
The number density of hard X-ray selected FSRQs has been found to
evolve strongly with redshift, with a peak number density
of \hbox{$z_{\rm peak}\approx 3$--4} (Ajello et al. 2009).
Unfortunately, the sample utilized by Ajello et~al. (2009) from the
\swift\ Burst Alert Telescope  
(BAT; Barthelmy et~al. 2005) does not have any FSRQ detections at $z>4$, and thus there is
considerable uncertainty in making quantitative comparisons at
the high redshifts most applicable to our work. We do note,
however, that the SED evolution we find for high-redshift HRLQs
(which are primarily FSRQs) will affect the interpretation of
number-density evolution results for X-ray selected samples.
For example, it may help to explain the higher $z_{\rm peak}$
value found for FSRQs than for general quasars or X-ray selected
AGNs. 

While the limited increase in \xray\ luminosities confirms earlier findings that 
the small-scale \xray\ jet-linked emission is not dominated by IC/CMB emission, 
our results are consistent with a fractional contribution from IC/CMB ($\sim6\%$ of the
nuclear \xray\ emission at $z=1.3$) rising with redshift. This does not conflict 
with growing evidence that RLQ jets have mildly relativistic bulk
velocities beyond several kpc, as inferred from radio core/jet
prominences (e.g., ${\gamma}_{\rm jet}\approx1.2-1.5$; Mullin \&
Hardcastle 2009), or that various predictions of 
IC/CMB models are not obviously met in \xray\ observations of large-scale RLQ
jets (e.g., Hardcastle~2006; Marshall et al.~2011; Massaro et al.~2011; note these
studies are based almost exclusively on RLQs with $z<2.5$). It may be that 
IC/CMB emission is relevant only over a limited range of spatial scales (e.g., 
$\approx1$--5~kpc), with quasar-related photon fields dominating at smaller distances 
(e.g., Ghisellini \& Tavecchio 2009) and the jet decelerating at
larger distances. For example, a drop in the bulk jet velocity from
$\beta=0.995$ to $\beta=0.75$, corresponding to a decrease from
${\gamma}=10$ at $\simlt$~5~kpc to ${\gamma}=1.5$ at large scales, would act to
diminish \hbox{X-ray} IC/CMB emission by a factor of $\approx 44$. Jet deceleration
appears to be required in at least some RLQs even in the context of
IC/CMB modeling of the large-scale \hbox{X-ray} jet emission, when
coupled with VLBI constraints (Hogan et al.~2011). Such 
deceleration could be partially due to mass entrainment; simulations
predict that even in powerful jets the bulk Lorentz factor decreases
by a factor of $\sim$2 on kpc scales (Bowman et al.~1996). Although
faster jets are more stable against Kelvin-Helmholtz instabilities
(e.g., Perucho et al.~2005), the gradual accumulation of
velocity-shear perturbations may similarly act to slow the jet. The
Compton-rocket effect (e.g., Ghisellini \& Tavecchio 2010) provides
another means of draining jets. Whatever the dominant mechanism, the
postulated degree of deceleration could plausibly be accomplished
without disrupting or decollimating the jet; for example, in their
classic work Kaiser \& Alexander (1997) estimate that a powerful jet
with ${\gamma}_{\rm jet}=2$ is persistently stable (although see also
Wang et al.~2011b). This deceleration possibility can be tested with
additional observations of large-scale jets in high-redshift
RLQs.\footnote{Cheung  
et al.~(2012) interpret the \xray/radio flux ratios of jets in 
GB~1428$+$4217 and GB~1508$+$4714 as supportive of an IC/CMB origin, but require a lower 
bulk Lorentz factor than is typically inferred from IC/CMB modeling of low-redshift RLQs.} 
In a fractional IC/CMB scenario, the enhanced \xray\ emission from
$z>4$ HRLQs is solely due to the increased energy density of the CMB;
the jet bulk velocities, the disk/corona accretion structure, and the
host-galaxy properties of individual RLQs are not required to evolve
with redshift.  

Alternative interpretations of the observed \xray\ emission enhancement of $z>4$ 
HRLQs are also possible. For example, increased host-galaxy star formation could provide additional
seed photons for external Compton upscattering in the jet (see $\S$4.1.3). 
Also, a decrease in the jet bulk Lorentz factor at high redshift (a possibility 
suggested by Volonteri et al.~2011) might increase the ratio of \xray\ to radio 
emission, particularly in a multi-component jet. It does not appear likely that 
\xray\ spectral curvature drives our results, as the observed-frame ultra-soft
luminosities are also enhanced. After careful consideration of selection effects, there
is no obvious indication that our sample contains a disproportionate number of ``extreme" 
HRLQs, with properties (low inclinations, strong variability) likely to correlate with 
enhanced \xray\ emission. 

Future \xray\ observations of additional $z>4$ HRLQs, 
e.g., a {\it Chandra\/} snapshot survey of the 11 objects listed in Table~2, would further
improve the sample statistics and better constrain the possible
mechanisms for the \xray\ enhancement. For example, the HRLQs at $z=3$--4
apparently have a similar level of \xray\ enhancement to that of \zfour\ HRLQs,
which would not agree with the expectations from the fractional IC/CMB
model. However, more objects are needed to reduce the large
uncertainties on the 
mean \daox\ values in current analyses. Identifying potential intrinsic \xray\ absorption 
in our objects, especially for the two moderately radio-loud quasars (J0741$+$2520 
and J1639$+$4340), requires deeper \xray\ spectroscopic observations. The 
ongoing {\it Fermi\/} LAT survey will provide tighter $\gamma$-ray luminosity upper 
limits or perhaps $\gamma$-ray detections of $z>4$ HRLQs, which will be useful for 
modeling their SEDs and for constraining the relative strengths of the
jet emissions from different mechanisms. 


\begin{acknowledgments}

We thank the anonymous referee for constructive comments. 
We thank M.~S.~Brotherton, A.~C.~Fabian, M.~J.~Hardcastle,
R.~F.~Mushotzky, E.~S.~Perlman, G.~T.~Richards, R.~W.~Romani,
and Z.~Shang for helpful discussions. 
We gratefully acknowledge the support of the ACIS Instrument Team contract \hbox{SV4-74018} 
(PI: G.~P.~Garmire) and NASA ADP grant NNX10AC99G (J.W., W.N.B). 
Funding for the SDSS and SDSS-II has been provided by the Alfred P. Sloan Foundation,
the Participating Institutions, the National Science Foundation, the U.S. Department
of Energy, the National Aeronautics and Space Administration, the Japanese Monbukagakusho,
the Max Planck Society, and the Higher Education Funding Council for England. The SDSS
Web site is http://www.sdss.org/. 

\end{acknowledgments}





\begin{center}
\begin{deluxetable}{llrrcllcccc}
\tablecolumns{10} \tabletypesize{\scriptsize}
\tablewidth{0pt}
\tablecaption{X-ray Observation Log\label{log_table}}
\tablehead{
    \colhead{} & \colhead{Object Name} & \colhead{R.A. (J2000)} & \colhead{Dec (J2000)} & \colhead{$z$\tablenotemark{a}} & \colhead{Detector} 
               & \colhead{Observation} & \colhead{Observation} & \colhead{Exp. Time\tablenotemark{b}} & \colhead{Reference\tablenotemark{c}} \\
    \colhead{} & \colhead{} & \colhead{(deg)} & \colhead{(deg)} & \colhead{}  & \colhead{}
               & \colhead{Date (UT)}  & \colhead{ID} & \colhead{(ks)} &
}
\startdata
\multicolumn{3}{l}{{\it Chandra} Cycle 12 Objects} & & & &\\
& SDSS~J$074154.71+252029.6$ & $115.4780$ & $25.3416$ & $5.194$ & ACIS-S & 2010 Oct 04 & 12171 & $~~4.0$ & 1,2 \\	
& SDSS~J$102623.61+254259.5$ & $156.5984$ & $25.7165$ & $5.304$ & ACIS-S & 2011 Mar 10 & 12167 & $~~5.0$ & 1 \\
& SDSS~J$141209.96+062406.9$ & $213.0415$ & $6.4019$ & $4.467$ & ACIS-S & 2011 Mar 16 & 12169 & $~~4.0$ & 1 \\
& SDSS~J$142048.01+120545.9$ & $215.2000$ & $12.0962$ & $4.027$ & ACIS-S & 2011 Mar 20 & 12168 & $~~4.0$ & 1 \\
& SDSS~J$163950.52+434003.6$ & $249.9605$ & $43.6677$ & $3.980$ & ACIS-S & 2010 Aug 08 & 12170 & $~~4.0$ & 1 \\
& SDSS~J$165913.23+210115.8$ & $254.8051$ & $21.0211$ & $4.784$ & ACIS-S & 2011 Jan 22 & 12172 & $~~6.5$ & 1 \\
\\
\multicolumn{3}{l}{Archival X-ray Data Objects} & & & & \\
& PSS~$0121+0347$\tablenotemark{d} & $20.3590$ & $3.7851$ & $4.130$ & ACIS-S & 2002 Feb 07 & 3151 & $~~5.7$  & 3 \\
& PMN~J$0324-2918$ & $51.1846$ & $-29.3059$ & $4.630$ & ACIS-S & 2003 Dec 16 & 4764 & $~~3.8$  & 4,5 \\
& PMN~J$0525-3343$\tablenotemark{e} & $81.2758$ & $-33.7182$ & $4.401$ & MOS\tablenotemark{f} & 2001 Sep 15 & 0050150301 & $19.0$  & 4,6,7 \\
& Q~$0906+6930$ & $136.6281$ & $69.5085$ & $5.480$ & ACIS-S & 2005 Jul 1 & 5637 & $30.0$  & 8,9 \\
& SDSS~J$091316.55+591921.6$\tablenotemark{d,{\rm g}} & $138.3190$ & $59.3227$ & $5.122$ & ACIS-S & 2002 Mar 7 & 3034 & $~~9.8$  & 10 \\
& RX~J$1028.6-0844$ & $157.1617$ & $-8.7441$ & $4.276$ & MOS\tablenotemark{f} & 2003 Jun 13 & 0153290101 & $21.3$ & 11,12,13 \\
& PMN~J$1155-3107$ & $178.7632$ & $-31.1330$ & $4.300$ & XRT & 2008 Jan 07 & 00036791002 & $~~4.2$ & 1,14 \\
& SDSS~J$123503.03-000331.7$\tablenotemark{d,{\rm g}} & $188.7626$ & $-0.0588$ & $4.673$ & MOS & 2010 Jul 01 & 0651740301 & $5.3$ & 1\\
& CLASS~J$1325+1123$\tablenotemark{d,{\rm g}} & $201.3021$ & $11.3916$ & $4.415$ & ACIS-S & 2003 Mar 02 & 3565 & $~~4.7$ & 15\\
& GB~$1428+4217$\tablenotemark{d,{\rm e}} & $217.5989$ & $42.0768$ & $4.715$ & MOS\tablenotemark{f} & 2003 Jan 17 & 0111260701 & $14.2$ & 16,17,18,19\\
& GB~$1508+5714$\tablenotemark{d,{\rm g}} & $227.5122$ & $57.0454$ & $4.313$ & ACIS-S & 2001 Jun 10 & 2241 & $90.1$ & 20,21,22,23\\
& GB~$1713+2148$\tablenotemark{d} & $258.8385$ & $21.7588$ & $4.011$ & ACIS-S & 2004 Jun 07 & 4815 & $~~9.5$ & 1,3,16\\
& PMN~J$1951+0134$ & $297.9001$ & $1.5785$ & $4.114$ & XRT & 2007 Mar 15 & 00036263002 & $~~7.8$ & 1,14\\
\enddata
\tablenotetext{a}{Redshift for each source. The redshift values for objects in the SDSS DR7 quasar catalog are from Hewett \& Wild (2010).}
\tablenotetext{b}{The exposure times are corrected for detector dead time.}
\tablenotetext{c}{References: (1) This work; (2) McGreer et~al. (2009); (3) Vignali et~al. (2003a); (4) Hook et~al. (2002); (5) Lopez et~al. (2006); (6) Fabian et~al. (2001); 
(7) Worsley et~al. (2004a); (8) Romani et~al. (2004); (9) Romani (2006); (10) Vignali et~al. (2003b); (11) Zickgraf et~al. (1997); (12) Grupe et~al. (2004); 
(13) Yuan et~al. (2005); (14) Healey et~al. (2008); (15) Bassett et~al. (2004); (16) Hook \& McMahon (1998); (17) Fabian et~al. (1998); 
(18) Fabian et~al. (1999); (19) Worsley et~al. (2004b); (20) Siemiginowska et~al. (2003); (21) Yuan et~al. (2003); (22) Cheung (2004); (23) Yuan et~al. (2006).}
\tablenotetext{d}{These archival sources are covered by the SDSS footprint.}
\tablenotetext{e}{These sources were observed in multiple \xmm\ or \swift\ observations. The parameters listed in this table are for the observation with the longest MOS or XRT exposure time. MOS exposure times have been corrected for the background flares.}
\tablenotetext{f}{These objects were observed by both the MOS and pn detectors. We list the MOS exposure times here.}
\tablenotetext{g}{These archival objects have SDSS spectroscopy; they appear in the SDSS DR7 quasar catalog.}
\end{deluxetable}
\end{center}

\begin{center}
\begin{deluxetable}{lrrcccrc}
\tablecolumns{8} \tabletypesize{\footnotesize}
\tablewidth{0pt}
\tablecaption{HRLQs at \zfour\ without Sensitive \xray\ Coverage\label{noxray_table}}
\tablehead{
    \colhead{Object Name} & \colhead{R.A. (J2000)} & \colhead{Dec (J2000)} & \colhead{$z$\tablenotemark{a}} & \colhead{$m_i$\tablenotemark{b}} 
               & \colhead{$M_i$} & \colhead{$f_{\rm 1.4~GHz}$\tablenotemark{c}} & \colhead{$\log R$} \\
    \colhead{} & \colhead{(deg)} & \colhead{(deg)} & \colhead{}  & \colhead{}
               & \colhead{}  & \colhead{(mJy)} & \colhead{} 
}
\startdata
SDSS~J$003126.79+150739.5$ & $7.8616$ & $15.1276$ & $4.297$ & $20.0$ & $-27.2$ & $41.0$ & $2.824$ \\ 
SDSS~J$030437.21+004653.5$ & $46.1550$ & $0.7815$ & $4.260$ & $20.2$ & $-27.1$ & $20.0$ & $3.259$ \\
SDSS~J$081333.32+350810.8$ & $123.3888$ & $35.1363$ & $4.945$ & $19.2$ & $-28.2$ & $20.0$ & $2.663$ \\
NVSS~J$105011-044254$ & $162.5480$ & $-4.7150$ & $4.270$ & $23.4$ & $-23.8$ & $~~6.0$ & $3.338$ \\
NVSS~J$112310-215405$ & $170.7921$ & $-21.9016$ & $4.110$ & $22.9$ & $-24.1$ & $48.0$ & $4.041$ \\
SDSS~J$123142.17+381658.9$ & $187.9257$ & $38.2830$ & $4.128$ & $20.1$ & $-26.9$ & $25.0$ & $2.777$ \\
SDSS~J$123726.26+651724.4$ & $189.3594$ & $65.2901$ & $4.300$ & $20.5$ & $-26.6$ & $24.0$ & $2.785$ \\
SDSS~J$124230.58+542257.3$ & $190.6274$ & $54.3826$ & $4.766$ & $19.7$ & $-27.7$ & $19.0$ & $2.678$ \\
PMN~J$2134-0419$ & $323.5501$ & $-4.3194$ & $4.350$ & $20.1$ & $-27.1$ & $333.0$ & $3.714$ \\
SDSS~J$222032.50+002537.5$ & $335.1354$ & $0.4271$ & $4.218$ & $19.9$ & $-27.1$ & $61.0$ & $3.531$ \\
PMN~J$2314+0201$ & $348.7030$ & $2.0309$ & $4.110$ & $20.2$ & $-26.9$ & $117.0$ & $3.308$ 
\enddata
\tablenotetext{a}{The redshift values for SDSS objects, all of which appear in the SDSS DR7 quasar catalog, are from Hewett \& Wild (2010).}
\tablenotetext{b}{The apparent $i$-band magnitude. For objects not in the SDSS footprint, the $i$-band magnitude was estimated from their 
$R$-band magnitude by assuming a power law with a spectral index of $\alpha_\nu=-0.5$.}
\tablenotetext{c}{Radio flux density at observed-frame 1.4~GHz from FIRST or NVSS.}
\end{deluxetable}
\end{center}

\clearpage
\input{ctstable}

\input{aoxtable}

\clearpage
\begin{deluxetable}{lcccccc}
\tabletypesize{\scriptsize}
\tablecaption{Results of Peto-Prentice Tests\label{twost_table}}
\tablewidth{0pt}
\tablehead{
\colhead{Parameter} & \colhead{No. of} & \colhead{No. of Comparison Objects} & \colhead{Mean} & \colhead{Mean} & \colhead{Statistic} & \colhead{Null-hypothesis} \\
\colhead{} & \colhead{Our HRLQs} & \colhead{(No. of \xray\ Limits)} & \colhead{(Our HRLQs)} & \colhead{(Comparison Objects)} & \colhead{} & \colhead{Probability}
}
\startdata
\daoxrq & $12$ & $283(14)$ & $0.435\pm0.054$ & $0.251\pm0.008$ & $3.844$ & $1.21\times10^{-4}$ \\
\daoxrl & $12$ & $283(14)$ & $0.167\pm0.049$ & $0.005\pm0.007$ & $3.814$ & $1.37\times10^{-4}$ \\
\\
\daoxrq\ (Worrall et~al.'s Objects) & $8$\tablenotemark{a} & $81(7)$ & $0.475\pm0.065$ & $0.302\pm0.018$ & $2.682$ & $7.32\times10^{-3}$ \\
\daoxrl\ (Worrall et al.'s Objects) & $8$ & $81(7)$ & $0.204\pm0.059$ & $0.014\pm0.016$ & $3.181$ & $1.47\times10^{-3}$ \\
\\
\daoxrq\ (FIRST-selected) & $8$ & $70(0)$ & $0.475\pm0.065$ & $0.241\pm0.011$ & $3.892$ & $9.94\times10^{-5}$ \\
\daoxrl\ (FIRST-selected) & $8$ & $70(0)$ & $0.205\pm0.059$ & $0.002\pm0.011$ & $3.814$ & $1.37\times10^{-4}$ \\
\\
\daoxrq\ ($\log L_{\rm 5~GHz} > 34$)\tablenotemark{b} & $12$ & $133(7)$ & $0.435\pm0.054$ & $0.294\pm0.012$ & $2.668$ & $7.63\times10^{-3}$ \\
\daoxrl\ ($\log L_{\rm 5~GHz} > 34$) & $12$ & $133(7)$ & $0.167\pm0.049$ & $-0.001\pm0.011$ & $3.628$ & $2.86\times10^{-4}$ \\
\\
\daoxrq\ ($\log L_{2500\mbox{\rm~\scriptsize\AA}} > 30.9$)\tablenotemark{b} & $12$ & $90(6)$ & $0.435\pm0.054$ & $0.305\pm0.016$ & $2.363$ & $1.81\times10^{-2}$ \\
\daoxrl\ ($\log L_{2500\mbox{\rm~\scriptsize\AA}} > 30.9$) & $12$ & $90(6)$ & $0.167\pm0.049$ & $0.018\pm0.013$ & $3.126$ & $1.77\times10^{-3}$ \\
\\
\daoxrq\ ($\alpha_{\rm r} > -0.5$) & $11$\tablenotemark{c} & $90(2)$ & $0.458\pm0.054$ & $0.261\pm0.014$ & $3.782$ & $1.56\times10^{-4}$ \\
\daoxrl\ ($\alpha_{\rm r} > -0.5$) & $11$ & $90(2)$ & $0.188\pm0.049$ & $0.029\pm0.013$ & $3.587$ & $3.35\times10^{-4}$ \\
\enddata
\tablecomments{For the detailed definition of the Peto-Prentice test statistic see Feigelson \& Nelson (1985). The null-hypothesis probability was calculated from the test statistic based on a Gaussian distribution, e.g., $1-P_G=1.21\times10^{-4}$, where $P_G$ is the cumulative Gaussian probability at $3.844\sigma$.}
\tablenotetext{a}{The number of radio-selected objects in our \zfour\ HRLQ sample that satisfy $m_i<20$. One object, GB~1713$+$2148, has $m_i>20$, and thus is excluded from the two-sample tests.}
\tablenotetext{b}{The range of radio luminosity $L_{\rm 5~GHz} > 34$ and optical/UV luminosity $L_{2500\mbox{\rm~\scriptsize\AA}} > 30.9$ were determined based on our \zfour\ HRLQs with $m_i<20$. Thus all 
12 of our HRLQs surviving this $m_i$ cut are included in these tests.}
\tablenotetext{c}{All of our 12 \zfour\ HRLQs with $m_i<20$ are FSRQs except for J1412$+$0624 which does not have available $\alpha_{\rm r}$ measurement.}
\end{deluxetable}

\begin{deluxetable}{cccc}
\tabletypesize{\footnotesize}
\tablecaption{Results of Peto-Prentice Tests for Additional Redshift Ranges\label{twostz_table}}
\tablewidth{0pt}
\tablehead{
\colhead{Sample} & \colhead{No. of HRLQs} & \colhead{Mean(\daoxrq)} & \colhead{Mean(\daoxrl)} \\
\colhead{} & \colhead{(No. of \xray\ Limits)} & \colhead{} & \colhead{} 
}
\startdata
$3\;\leqslant\;z\;<\;4$ & $9(1)$ & $0.468\pm0.056$ & $0.167\pm0.046$ \\
$z\;<\;3$ & $274(13)$  & $0.244\pm0.008$ & $0.000\pm0.007$  \\
\hline
Test Statistic & & $4.954$ & $5.428$ \\
Null-hypothesis Probability & & $7.27\times10^{-7}$ & $5.70\times10^{-8}$\\
\hline\hline\\
$2\;\leqslant\;z\;<\;3$ & $39(4)$ & $0.249\pm0.023$ & $-0.030\pm0.019$ \\
$z\;<\;2$ & $235(9)$  & $0.243\pm0.009$ & $0.005\pm0.008$  \\
\hline
Test Statistic & & $0.176$ & $1.981$ \\
Null-hypothesis Probability & & $0.861$ & $0.048$\\
\hline\hline\\
$z\;\geqslant\;3$ & $21(1)$ & $0.450\pm0.039$ & $0.167\pm0.034$ \\
$z\;<\;3$ & $274(13)$  & $0.244\pm0.008$ & $0.000\pm0.007$ \\
\hline
Test Statistic & & $5.905$ &$5.726$ \\
Null-hypothesis Probability & & $3.53\times10^{-9}$ & $1.03\times10^{-8}$\\
\enddata
\tablecomments{For the detailed definition of the Peto-Prentice test
  statistic see Feigelson \& Nelson (1985). The null-hypothesis
  probability was calculated from the test statistic based on a
  Gaussian distribution, e.g., $1-P_G=7.27\times10^{-7}$, where $P_G$
  is the cumulative Gaussian probability at $4.954\sigma$.} 
\end{deluxetable}

\clearpage
\begin{center}
\begin{deluxetable}{clccccccc}
\tabletypesize{\scriptsize}
\tablecaption{X-ray Properties of \zfour\ HRLQs Derived from the Observed-Frame Ultrasoft Band (0.3--1.0~keV)\label{us_table}}
\tablewidth{0pt}
\tablehead{ \colhead{} & \colhead{Object Name} & \colhead{Count Rate\tablenotemark{a}} & \colhead{$F_{\rm X, US}$\tablenotemark{b}} & \colhead{$f_{\rm 2~keV, US}$\tablenotemark{c}}
            & \colhead{$\alpha_{\rm ox, US}$\tablenotemark{d}} & \colhead{$\Delta\alpha_{\rm ox, RQQ, US}$\tablenotemark{e}} & \colhead{$\Delta\alpha_{\rm ox, RLQ, US}$\tablenotemark{f}} 
            & \colhead{$\alpha_{\rm ox, US}-\alpha_{\rm ox}$\tablenotemark{g}}
}
\startdata
\multicolumn{3}{l}{{\it Chandra} Cycle 12 Objects} & & & \\
& SDSS~J$141209.96+062406.9$ & $1.32^{+0.86}_{-0.55}$ & $0.81$ & $5.66$ & $-1.56$ & $0.13$ & $-0.11$ & $-0.05$ \\
& SDSS~J$142048.01+120545.9$ & $3.89^{+1.26}_{-0.97}$ & $2.48$ & $18.75$ & $-1.33$ & $0.34$ & $0.07$ & $0.01$ \\
\\
\multicolumn{3}{l}{Archival X-ray Data Objects} & & &\\
& PSS~$0121+0347$ & $5.93^{+1.21}_{-1.02}$ & $2.02$ & $18.41$ & $-1.53$ & $0.22$ & $-0.02$ & $-0.06$ \\
& PMN~J$0324-2918$ & $6.62^{+1.66}_{-1.35}$ & $4.37$ & $38.69$ & $-1.40$ & $0.35$ & $0.08$ & $0.00$ \\
& PMN~J$0525-3343$ & $34.33^{+0.87}_{-0.85}$ & $22.07$ & $178.82$ & $-1.15$ & $0.60$ & $0.33$ & $0.02$ \\
& Q~$0906+6930$ & $5.32^{+0.47}_{-0.43}$ & $3.55$ & $31.09$ & $-1.26$ & $0.45$ & $0.18$ & $0.05$ \\
& RX~J$1028.6-0844$ & $21.80^{+0.73}_{-0.70}$ & $16.25$ & $118.41$ & $-1.16$ & $0.56$ & $0.27$ & $-0.07$ \\
& CLASS~J$1325+1123$ & $2.01^{+0.89}_{-0.64}$ & $0.75$ & $6.43$ & $-1.61$ & $0.11$ & $-0.13$ & $-0.08$ \\
& GB~$1428+4217$ & $53.55^{+1.41}_{-1.37}$ & $32.45$ & $281.56$ & $-1.00$ & $0.73$ & $0.45$ & $-0.07$ \\
& GB~$1508+5714$ & $22.33^{+0.51}_{-0.50}$ & $6.37$ & $48.61$ & $-1.03$ & $0.60$ & $0.27$ & $-0.07$ \\
\enddata
\tablecomments{The objects included in this table are our \zfour\ HRLQs with $m_i<20$, i.e., those \zfour\ HRLQs used in 
the two-sample analyses in \S\ref{discuss:daox}, except for the two \swift\ sources (PMN~J1155$-$3107 and PMN~J1951$+$0134).}
\tablenotetext{a}{The \chandra\ or \xmm\ count rate in the observed-frame ultrasoft \xray\ band (0.3--1.0 keV), in units of $10^{-3}$~s$^{-1}$.}
\tablenotetext{b}{The Galactic-absorption corrected flux in the observed-frame ultrasoft \xray\ band, in units of $10^{-14}$~erg~cm$^{-2}$~s$^{-1}$.}
\tablenotetext{c}{The \xray\ flux density at rest-frame 2~keV derived from the ultrasoft band count rate, in units of $10^{-32}$~erg~cm$^{-2}$~s$^{-1}$~Hz$^{-1}$.}
\tablenotetext{d}{The \aox\ values obtained from the ultrasoft \xray\ band data.}
\tablenotetext{e}{The \daoxrq\ values obtained from the ultrasoft \xray\ band data.}
\tablenotetext{f}{The \daoxrl\ values obtained from the ultrasoft \xray\ band data.}
\tablenotetext{g}{The difference between the $\alpha_{\rm ox, US}$ values in this table and the \aox\ values in Column (15) of Table~\ref{aox_table} which are obtained from 
the observed-frame soft band (0.5--2.0~keV) data. The \xray\ spectral curvature discussed in \S\ref{discuss:daox:physical} is expected to result in negative values for this difference.}
\end{deluxetable}
\end{center}

\begin{center}
\begin{deluxetable}{clccc}
\tabletypesize{\footnotesize}
\tablecaption{Emission-Line REW Measurements\label{ew_table}}
\tablewidth{0pt}
\tablehead{ \colhead{} & \colhead{Object Name} & \colhead{MJD\tablenotemark{a}} & \colhead{REW(\lyanv)} & \colhead{REW(\civ)}
}
\startdata
\multicolumn{3}{l}{{\it Chandra} Cycle 12 Objects} & &\\
& SDSS~J$102623.61+254259.5$\tablenotemark{b} & $53734$ & $41.6$ & \nodata \\
& SDSS~J$141209.96+062406.9$ & $53504$ & $2.4$ & $7.1$ \\
& SDSS~J$142048.01+120545.9$ & $53885$ & $55.0$ & $52.7$ \\
& SDSS~J$163950.52+434003.6$ & $52051$ & $23.3$ & $7.8$ \\
& SDSS~J$165913.23+210115.8$ & $52913$ & $17.9$ & $6.4$ \\
\\
\multicolumn{3}{l}{Archival X-ray Data Objects} & &\\
& PSS~$0121+0347$\tablenotemark{c} & $53328$ & $81.1$ & $33.5$ \\
& SDSS~J$091316.55+591921.6$\tablenotemark{b} & $51907$ & $110.9$ & \nodata \\
& SDSS~J$123503.03-000331.7$ & $51941$ & $39.5$ & $54.1$ \\
& CLASS~J$1325+1123$ & $53148$ & $75.8$ & $63.4$ \\
& GB~$1428+4217$\tablenotemark{d} & $50283$ & $17.1$ & $12.9$ \\
& GB~$1508+5714$ & $52079$ & $60.1$ & $47.0$\\
& GB~$1713+2148$\tablenotemark{d} & $50280$ & $107.0$ & $127.7$ \\
\enddata
\tablecomments{All REW values are in units of \AA.}
\tablenotetext{a}{The MJD (Modified Julian Date) listed here is the date of optical/UV spectroscopy.}
\tablenotetext{b}{Two objects (J0913$+$5919 and J1026$+$2542) do not have \civ\ coverage in their SDSS spectra because of their high redshifts.}
\tablenotetext{c}{Reference for emission-line REW measurements: Vignali et~al. (2003a).}
\tablenotetext{d}{Reference for emission-line REW measurements: Hook \& McMahon (1998).}
\end{deluxetable}
\end{center}

\clearpage
\begin{center}
\begin{deluxetable}{clcccc}
\tabletypesize{\footnotesize}
\tablecaption{Non-Thermal Dominance (NTD) Calculation\label{ntd_table}}
\tablewidth{0pt}
\tablehead{ \colhead{} & \colhead{Object Name} & \colhead{$\log L_{\rm C~IV}$} & \colhead{$\log \nu L_{\rm 1350, obs}$} & \colhead{$\log \nu L_{\rm 1350, pred}$} & \colhead{NTD}\\
            \colhead{} & \colhead{} & \colhead{(erg s$^{-1}$)} & \colhead{(erg s$^{-1}$)} & \colhead{(erg s$^{-1}$)} & \colhead{}
}
\startdata
\multicolumn{3}{l}{{\it Chandra} Cycle 12 Objects} & &\\
& SDSS~J$141209.96+062406.9$ & $44.289$ & $46.670$ & $45.881$ & $6.150$\\
& SDSS~J$142048.01+120545.9$ & $44.968$ & $46.465$ & $46.467$ & $0.995$\\
& SDSS~J$163950.52+434003.6$ & $44.946$ & $47.255$ & $46.448$ & $6.402$\\
& SDSS~J$165913.23+210115.8$ & $44.121$ & $46.516$ & $45.736$ & $6.021$\\
\\
\multicolumn{3}{l}{Archival X-ray Data Objects} & &\\
& SDSS~J$123503.03-000331.7$ & $44.597$ & $46.100$ & $46.148$ & $0.896$ \\
& CLASS~J$1325+1123$ & $45.425$ & $46.860$ & $46.861$ & $0.997$\\
& GB~$1508+5714$ & $44.873$ & $46.503$ & $46.385$ & $1.312$ \\
\enddata
\end{deluxetable}
\end{center}


\begin{center}
\begin{deluxetable}{ccrrrrrrrrr}
\tabletypesize{\footnotesize}
\tablecaption{Correlation Analysis Results\label{corr_table}}
\tablewidth{0pt}
\tablehead{
\colhead{} & & \multicolumn{4}{c}{REW(\lyanv)} & & \multicolumn{4}{c}{REW(\civ)} \\
     \cline{3-6}\cline{8-11}
     \colhead{} & & $N$\tablenotemark{a} & &\multicolumn{2}{c}{Spearman} & & $N$\tablenotemark{a} & & \multicolumn{2}{c}{Spearman} \\
     \cline{5-6}\cline{10-11}
     & & & & \colhead{$r_S$} & \colhead{$1-P_S$} & & & & \colhead{$r_S$} & \colhead{$1-P_S$} 
}
\startdata
\daoxrq & & $32$ & & $-0.19$ & $71.0\%$ & & $169$ & & $0.14$ & $93.3\%$ \\
\daoxrl & & $32$ & & $-0.21$ & $75.0\%$ & & $169$ & & $0.08$ & $69.0\%$ \\
$\log R$ & & $298$ & & $0.14$ & $98.0\%$ & & $3536$ & & $0.30$ & $>99.99\%$ \\
\enddata
\tablenotetext{a}{$N$ is the sample size.}
\end{deluxetable}
\end{center}

\begin{deluxetable}{cccccccc}
\tablecolumns{8} \tabletypesize{\footnotesize}
\tablewidth{0pt}
\tablecaption{Joint X-ray Spectral Analysis Results}
\tablehead{
\colhead{}
    & \colhead{}
    & \multicolumn{2}{c}{Power Law}
    & \colhead{}
    & \multicolumn{3}{c}{Power Law}\\
    \colhead{}
    & \colhead{}
    & \multicolumn{2}{c}{with Galactic Absorption}
    & \colhead{}
    & \multicolumn{3}{c}{with Galactic and Intrinsic Absorption}
    \\
    \cline{3-4}\cline{6-8}\\    
\colhead{Object Name} 
    & \colhead{}
    & \colhead{$\Gamma_{\rm X}$} 
    & \colhead{$C/n$\tablenotemark{a}} 
    & \colhead{}
    & \colhead{$\Gamma_{\rm X}$} 
    & \colhead{$N_H (10^{22} \ {\rm cm}^{-2})$}
    & \colhead{$C/n$\tablenotemark{a}}
}
\startdata
J1026/J1412/J1420/J1659\tablenotemark{b} & & $1.59^{+0.24}_{-0.24}$ & $122.47/148$ & & $1.60^{+0.40}_{-0.24}$ & $<9.50$ & $122.47/148$ \\
J0741/J1659\tablenotemark{c} & & $1.34^{+0.39}_{-0.37}$ & $46.87/66$ & & $1.34^{+0.49}_{-0.36}$ & $<12.41$ & $46.87/66$ \\
\enddata
\tablenotetext{a}{$C$ is the $C$-statistic, while $n$ is the total number of spectral bins.}
\tablenotetext{b}{Our \chandra\ Cycle~12 targets with $\log R>2.5$. The names are in the format of 'J$hhmm$' for brevity.} 
\tablenotetext{c}{Our \chandra\ Cycle~12 targets that are moderately radio loud. The names are in the format of 'J$hhmm$' for brevity.}
\label{xspec_table}
\end{deluxetable}

\clearpage
\begin{center}
\begin{deluxetable}{llcccc}
\tablecolumns{6} \tabletypesize{\scriptsize}
\tablewidth{0pt}
\tablecaption{Upper Limits upon \garay\ Flux and Luminosity from the {\it Fermi} LAT\label{fermi_table}}
\tablehead{
    \colhead{} & \colhead{Object Name} & \colhead{Photon Flux Limit\tablenotemark{a}} & \colhead{Energy Flux Limit\tablenotemark{b}} & \colhead{Flux Density Limit\tablenotemark{c}} & \colhead{$\log(\nu L_{\rm 1~GeV}$) Limit\tablenotemark{d}} \\
    \colhead{} & \colhead{} & \colhead{($10^{-9}$~photons~cm$^{-2}$~s$^{-1}$)} & \colhead{($10^{-12}$~erg~cm$^{-2}$~s$^{-1}$)} & \colhead{($10^{-35}$~erg~cm$^{-2}$~s$^{-1}$~Hz$^{-1}$)} 
    & \colhead{(erg~s$^{-1}$~Hz$^{-1}$)}
}
\startdata
\multicolumn{3}{l}{{\it Chandra} Cycle 12 Objects} & & \\
& SDSS~J$074154.71+252029.6$ & $4.5$ & $2.4$ & $2.1$ & $47.4$ \\
& SDSS~J$102623.61+254259.5$ & $4.5$ & $2.4$ & $2.2$ & $47.4$ \\
& SDSS~J$141209.96+062406.9$ & $5.0$ & $2.6$ & $2.0$ & $47.3$ \\
& SDSS~J$142048.01+120545.9$ & $5.0$ & $2.6$ & $1.8$ & $47.1$ \\
& SDSS~J$163950.52+434003.6$ & $4.5$ & $2.4$ & $1.6$ & $47.1$ \\
& SDSS~J$165913.23+210115.8$ & $4.5$ & $2.4$ & $1.9$ & $47.3$ \\
\\
\multicolumn{3}{l}{Archival X-ray Data Objects} & &  \\
& PSS~$0121+0347$ & $4.5$ & $2.4$ & $1.6$ & $47.1$ \\
& PMN~J$0324-2918$ & $4.5$ & $2.4$ & $1.9$ & $47.3$ \\
& PMN~J$0525-3343$ & $4.5$ & $2.4$ & $1.8$ & $47.2$ \\
& Q~$0906+6930$ & $3.5$ & $1.8$ & $1.8$ & $47.4$ \\
& SDSS~J$091316.55+591921.6$ & $4.0$ & $2.1$ & $1.9$ & $47.3$ \\
& RX~J$1028.6-0844$ & $5.0$ & $2.6$ & $1.9$ & $47.2$ \\
& PMN~J$1155-3107$ & $5.0$ & $2.6$ & $1.9$ & $47.2$ \\
& SDSS~J$123503.03-000331.7$ & $5.0$ & $2.6$ & $2.1$ & $47.3$ \\
& CLASS~J$1325+1123$ & $5.0$ & $2.6$ & $2.0$ & $47.3$ \\
& GB~$1428+4217$ & $4.0$ & $2.1$ & $1.7$ & $47.2$ \\
& GB~$1508+5714$ & $4.0$ & $2.1$ & $1.5$ & $47.1$ \\
& GB~$1713+2148$ & $5.0$ & $2.6$ & $1.8$ & $47.1$ \\
& PMN~J$1951+0134$ & $7.0$ & $3.7$ & $2.5$ & $47.3$ \\
\enddata
\tablenotetext{a}{The upper limit upon the {\it Fermi} LAT photon flux between observed-frame 100~MeV--100~GeV.}
\tablenotetext{b}{The upper limit upon the \garay\ energy flux between observed-frame 100~MeV--100~GeV.}
\tablenotetext{c}{The upper limit upon the \garay\ flux density at rest-frame 1~GeV.}
\tablenotetext{d}{The upper limit upon the monochromatic \garay\ luminosity at rest-frame 1~GeV.}
\end{deluxetable}
\end{center}

\begin{center}
\begin{deluxetable}{clccc}
\tabletypesize{\footnotesize}
\tablecaption{Measurements of Black-Hole Mass and Eddington Ratio\label{mbh_table}}
\tablewidth{0pt}
\tablehead{ \colhead{} & \colhead{Object Name} & &\colhead{$\log M_{\rm BH}$ ($M_\odot$)} & \colhead{$\log (L/L_{\rm Edd})$}
}
\startdata
\multicolumn{3}{l}{{\it Chandra} Cycle 12 Objects}& &\\
& SDSS~J$141209.96+062406.9$ & & $~~9.859\pm1.218$ & $-0.708$ \\
& SDSS~J$142048.01+120545.9$ & & $~~9.284\pm0.180$ & $-0.338$ \\
& SDSS~J$163950.52+434003.6$ & & $10.628\pm0.168$ & $-0.893$ \\
\\
\multicolumn{3}{l}{Archival X-ray Data Objects} & &\\
& SDSS~J$123503.03-000331.7$ & & $~~9.182\pm0.727$ & $-0.602$ \\
& CLASS~J$1325+1123$ & & $~~9.455\pm0.071$ & $-0.115$ \\
& GB~$1508+5714$ & & $~~8.517\pm0.388$ & $~~~0.467$\\
\enddata
\end{deluxetable}
\end{center}


\clearpage

\begin{figure*}[t]
    \centering
    \includegraphics[width=5.3in]{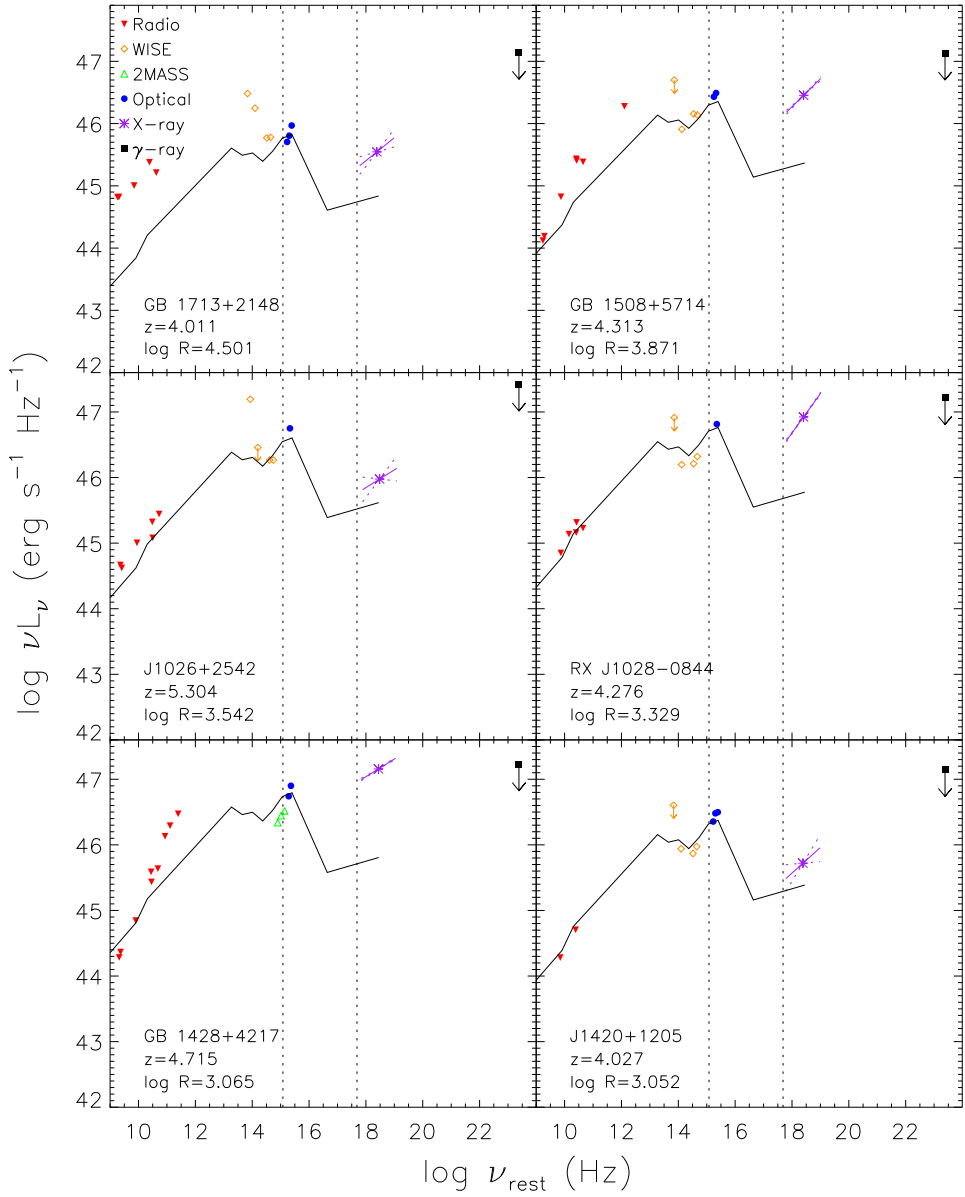}
    \caption{\footnotesize{The rest-frame broad-band SEDs for our sample of objects ordered by $\log R$ (in descending order), including radio 
     (red filled upside-down triangles), mid-infrared (\wise, orange open diamonds), near-infrared (2MASS and UKIRT, 
     green open triangles), optical (blue filled circles), \xray\ (purple asterisks and lines), and \garay\ (filled square) data points. Downward 
     arrows indicate upper limits. The purple solid (dotted) lines show the \xray\ power-law spectra (and their uncertainty 
     range) based on the photon-index values provided in Column 9 of Table~\ref{aox_table}. The purple asterisks represent 
     observed-frame 2~keV. The black solid lines show the composite
     SEDs for the 10 RLQs in Shang et~al. (2011; S11) with comparable
     optical luminosity and radio loudness, scaled 
     to the flux density at rest-frame 2500~\AA\ (corresponding to
     $10^{15.1}$~Hz; see \S\ref{discuss:sed}). The vertical dotted lines show
     the frequencies of rest-frame 2500~\AA\ and 2~keV. The name, redshift, and $\log R$ of each object are labeled at the bottom right of each panel.}}
    \label{seds_fig}
\end{figure*}

\begin{figure*}[t]
    \figurenum{\ref{seds_fig}}
    \centering
    \includegraphics[width=5.8in]{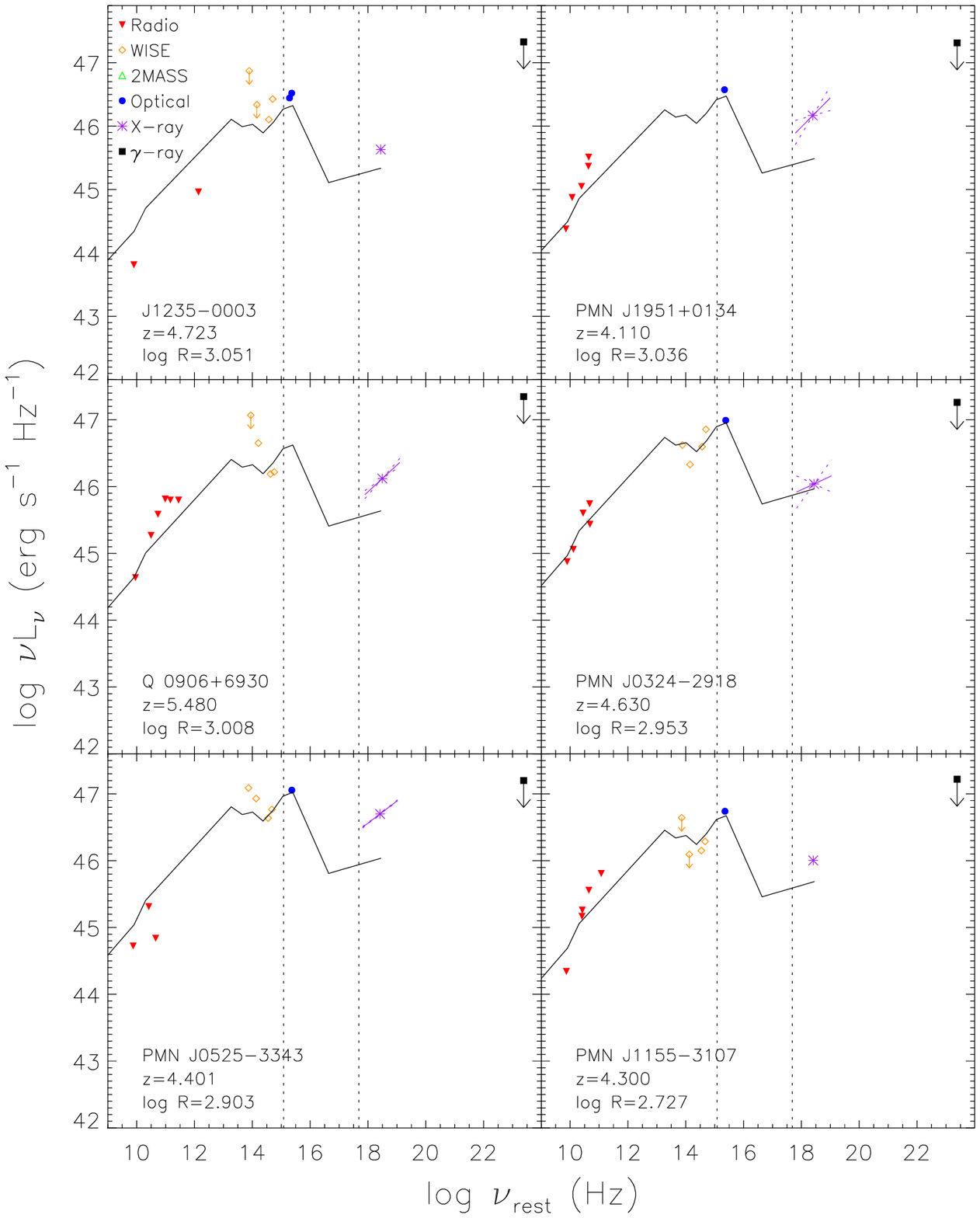}
    \caption{\footnotesize{\it Continued.}}
\end{figure*}

\begin{figure*}[t]
    \figurenum{\ref{seds_fig}}
    \centering
    \includegraphics[width=5.3in]{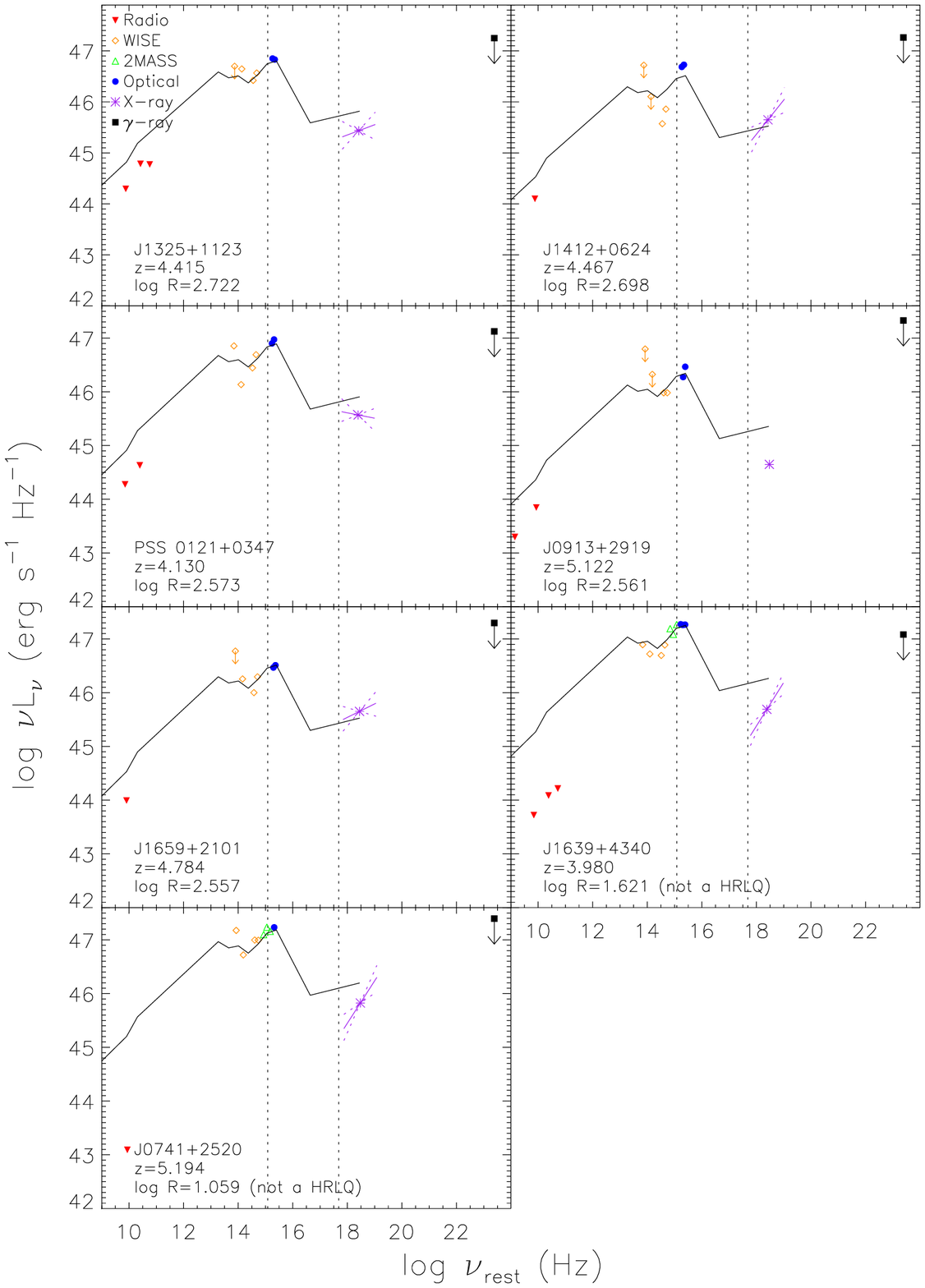}
    \caption{\footnotesize{\it Continued.}}
\end{figure*}




\end{document}

%% file: ctstable.tex
\begin{center}
\begin{deluxetable}{llccccc}
\tablecolumns{7} \tabletypesize{\footnotesize}
\tablewidth{0pt}
\tablecaption{X-ray Counts for Objects without Previously Published X-ray Photometry\label{cts_table}}
\tablehead{
    \colhead{}     
    & \colhead{}                     
    & \colhead{Full Band}     
    & \colhead{Soft Band}     
    & \colhead{Hard Band}     
    & \colhead{Band}
    & \colhead{}\\
    \colhead{}
    & \colhead{Object Name} 
    & \colhead{(0.5--8.0 keV)\tablenotemark{a}}
    & \colhead{(0.5--2.0 keV)\tablenotemark{a}} 
    & \colhead{(2.0--8.0 keV)\tablenotemark{a}} 
    & \colhead{Ratio\tablenotemark{b}}  
    & \colhead{$\Gamma$\tablenotemark{c}}
}
\startdata
\multicolumn{2}{l}{{\it Chandra} Cycle 12 Objects} & & & \\
& SDSS~J$074154.71+252029.6$ & $29.8^{+6.5}_{-5.4}$ & $17.8^{+5.3}_{-4.2}$ & $11.1^{+4.4}_{-3.3}$ & $0.62^{+0.31}_{-0.23} $ & $1.21^{+0.44}_{-0.37} $ \\
& SDSS~J$102623.61+254259.5$ & $59.3^{+8.7}_{-7.7}$ & $42.6^{+7.6}_{-6.5}$ & $14.3^{+4.9}_{-3.7}$ & $0.34^{+0.13}_{-0.10} $ & $1.73^{+0.32}_{-0.31} $ \\
& SDSS~J$141209.96+062406.9$ & $29.9^{+6.5}_{-5.4}$ & $18.9^{+5.4}_{-4.3}$ & $9.9^{+4.3}_{-3.1}$ & $0.52^{+0.27}_{-0.20} $ & $1.33^{+0.43}_{-0.38} $ \\
& SDSS~J$142048.01+120545.9$ & $48.9^{+8.0}_{-7.0}$ & $33.9^{+6.9}_{-5.8}$ & $13.1^{+4.7}_{-3.6}$ & $0.39^{+0.16}_{-0.12} $ & $1.61^{+0.35}_{-0.32} $ \\
& SDSS~J$163950.52+434003.6$ & $42.9^{+7.6}_{-6.5}$ & $25.8^{+6.2}_{-5.0}$ & $15.7^{+5.1}_{-3.9}$ & $0.61^{+0.25}_{-0.19} $ & $1.19^{+0.34}_{-0.31} $ \\
& SDSS~J$165913.23+210115.8$ & $43.6^{+7.7}_{-6.6}$ & $30.8^{+6.6}_{-5.5}$ & $11.1^{+4.4}_{-3.3}$ & $0.36^{+0.16}_{-0.12} $ & $1.75^{+0.40}_{-0.36} $ \\
\\
\multicolumn{3}{l}{Archival X-ray Data Objects} & & & & \\
& PMN~J$1155-3107$ & $12.1^{+4.6}_{-3.4}$ & $10.1^{+4.3}_{-3.1}$ & $<8.3$ & $<0.82$ & $>1.33$ \\ 
& SDSS~J$123503.03-000331.7$ & $25.9^{+9.1}_{-8.1}$ & $<18.3$ & $<15.9$ & \nodata & \nodata \\
& GB~$1713+2148$ & $92.5^{+10.7}_{-9.6}$ & $63.2^{+9.0}_{-7.9}$ & $25.9^{+6.2}_{-5.0}$ & $0.41^{+0.11}_{-0.09} $ & $1.63^{+0.24}_{-0.23} $ \\
& PMN~J$1951+0134$ & $45.2^{+7.8}_{-6.7}$ & $25.4^{+6.1}_{-5.0}$ & $20.7^{+5.6}_{-4.5}$ & $0.81^{+0.30}_{-0.24}$ & $1.54^{+0.33}_{-0.30}$ \\
\enddata
\tablenotetext{a}{Errors on the \hbox{X-ray} counts were calculated using Poisson statistics corresponding to the 1$\sigma$ significance level according to Tables 1 and 2 of Gehrels (1986).}
\tablenotetext{b}{The band ratio is defined here as the number of hard-band counts divided by the number of soft-band counts. The errors on the band ratio correspond to the 1$\sigma$ significance level and were calculated using equation (1.31) in \S 1.7.3 of Lyons (1991). The band ratios for all of the {\it Chandra} objects observed in the same cycle can be directly compared with one another.}
\tablenotetext{c}{The effective power-law photon indices were calculated using the PIMMS tool (version 3.9$k$). The effects of the quantum-efficiency decay over time at low energies of the ACIS detector were corrected for {\it Chandra} observed objects. The \chandra\ ACIS Cycle~5 response was used for GB~1713+2148, while the \chandra\ Cycle~12 response was used for the other objects. For J1155$-$3107 and J1235$-$0003 which do not have photon index estimation, we adopt a typical RLQ photon index value $\Gamma=1.6$ in the following analyses.}
\end{deluxetable}
\end{center}

%% file: aoxtable.tex
\begin{turnpage}
\begin{deluxetable}{lccccrrcccccccccc}
\tablecolumns{17} \tabletypesize{\tiny}
\tablewidth{0pt}
\tablecaption{X-ray, Optical/UV, and Radio Properties\label{aox_table}}
\tablehead{
     \colhead{}
    & \colhead{}        
    & \colhead{}        
    & \colhead{}            
    & \colhead{Count}
    & \colhead{}                       
    & \colhead{}                 
    & \colhead{log $L_{\rm x}$}                 
    & \colhead{}
    & \colhead{}
    & \colhead{log $L_{\rm uv}$}          
    & \colhead{}
    & \colhead{log $L_{\rm r}$}
    & \colhead{} 
    & \colhead{} 
    & \colhead{} 
    & \colhead{} \\
     \colhead{Object Name} 
    & \colhead{$m_{i}$\tablenotemark{a}} 
    & \colhead{$M_{i}$\tablenotemark{b}} 
    & \colhead{$N_{\rm H}$} 
    & \colhead{Rate\tablenotemark{c}} 
    & \colhead{$F_{\rm X}$\tablenotemark{d}} 
    & \colhead{$f_{\rm 2\;keV}$\tablenotemark{e}} 
    & \colhead{($2-10\;{\rm keV}$)\tablenotemark{f}} 
    & \colhead{$\Gamma_{\rm X}$\tablenotemark{g}}    
    & \colhead{$f_{2500\mbox{\rm~\scriptsize\AA}}$\tablenotemark{h}} 
    & \colhead{(2500 \AA)\tablenotemark{i}}  
    & \colhead{$\alpha_{\rm r}$\tablenotemark{j}}
    & \colhead{(5~GHz)\tablenotemark{k}}
    & \colhead{$\log R$\tablenotemark{l}}     
    & \colhead{$\alpha_{\rm ox}$} 
    & \colhead{$\Delta \alpha_{\rm ox, RQQ}$ $(\sigma)$\tablenotemark{m}}
    & \colhead{$\Delta \alpha_{\rm ox, RLQ}$\tablenotemark{n}} \\
     \colhead{(1)} 
    & \colhead{(2)} 
    & \colhead{(3)} 
    & \colhead{(4)} 
    & \colhead{(5)} 
    & \colhead{(6)} 
    & \colhead{(7)} 
    & \colhead{(8)} 
    & \colhead{(9)} 
    & \colhead{(10)} 
    & \colhead{(11)} 
    & \colhead{(12)} 
    & \colhead{(13)} 
    & \colhead{(14)}
    & \colhead{(15)} 
    & \colhead{(16)} 
    & \colhead{(17)}
}
\startdata
\multicolumn{2}{l}{{\it Chandra} Cycle 12 Objects} & & & \\
SDSS~J$074154.71+252029.6$ & $18.54$ & $-29.04$ & $4.26$ & $4.44^{+1.33}_{-1.04}$ & $1.89$ & $6.80$ & $45.20$ & $1.21^{+0.44}_{-0.37}$ & $1.95$ & $31.97$ & \nodata & $33.16$ & $1.059$ & $-1.71$ & $0.07$ $(0.5)$ & $-0.02$  \\
SDSS~J$102623.61+254259.5$ & $20.03$ & $-27.50$ & $1.80$ & $8.55^{+1.52}_{-1.30}$ & $3.51$ & $24.03$ & $45.76$ & $1.73^{+0.32}_{-0.31}$ & $0.62$ & $31.49$ & $-0.38$ & $35.15$ & $3.542$ & $-1.31$ & $0.39$ $(2.7)$ & $0.07$ \\
SDSS~J$141209.96+062406.9$ & $19.44$ & $-27.74$ & $2.11$ & $4.73^{+1.36}_{-1.08}$ & $1.93$ & $7.73$ & $45.15$ & $1.33^{+0.43}_{-0.38}$ & $0.65$ & $31.40$ & \nodata & $34.36$ & $2.698$ & $-1.51$ & $0.18$ $(1.2)$ & $-0.06$  \\
SDSS~J$142048.01+120545.9$ & $19.80$ & $-27.18$ & $1.72$ & $8.47^{+1.72}_{-1.45}$ & $3.45$ & $17.84$ & $45.44$ & $1.61^{+0.35}_{-0.32}$ & $0.56$ & $31.26$ & $-0.36$ & $34.55$ & $3.052$ & $-1.34$ & $0.33$ $(2.3)$ & $0.06$ \\
SDSS~J$163950.52+434003.6$ & $17.78$ & $-29.13$ & $1.36$ & $6.45^{+1.54}_{-1.26}$ & $2.57$ & $8.67$ & $45.13$ & $1.19^{+0.34}_{-0.31}$ & $4.29$ & $32.14$ & $-0.32$ & $33.93$ & $1.621$ & $-1.80$ & $-0.01$ $(0.1)$ & $-0.16$ \\
SDSS~J$165913.23+210115.8$ & $20.26$ & $-27.17$ & $5.47$ & $4.77^{+1.02}_{-0.85}$ & $2.15$ & $14.12$ & $45.34$ & $1.75^{+0.40}_{-0.36}$ & $0.60$ & $31.40$ & \nodata & $34.25$ & $2.557$ & $-1.39$ & $0.30$ $(2.1)$ & $0.07$  \\
\\
\multicolumn{2}{l}{Archival \xray\ Data Objects} & & & \\
PSS~$0121+0347$ & $18.57$ & $-28.44$ & $3.19$ & $10.30^{+1.60}_{-1.30}$ & $3.19$ & $26.78$ & $45.64$ & $2.10^{+0.40}_{-0.30}$ & $1.77$ & $31.78$ & $-0.33$ & $34.47$ & $2.573$ & $-1.47$ & $0.28$ $(1.9)$ & $0.04$ \\
PMN~J$0324-2918$ & $18.65$ & $-28.61$ & $1.19$ & $13.70^{+2.20}_{-1.90}$ & $5.83$ & $39.70$ & $45.89$ & $1.80^{+0.40}_{-0.40}$ & $1.73$ & $31.84$ & $+0.30$ & $34.92$ & $2.953$ & $-1.40$ & $0.35$ $(2.4)$ & $0.08$ \\
PMN~J$0525-3343$ & $18.63$ & $-28.52$ & $2.19$ & \nodata & $27.70$ & $159.40$ & $46.46$ & $1.67^{+0.02}_{-0.02}$ & $1.73$ & $31.81$ & $+0.06$ & $34.84$ & $2.903$ & $-1.17$ & $0.58$ $(4.0)$ & $0.31$ \\
Q~$0906+6930$ & $19.85$ & $-27.76$ & $3.64$ & $10.72^{+0.63}_{-0.60}$ & $4.16$ & $24.78$ & $45.80$ & $1.6^{+0.1}_{-0.1}$ & $0.61$ & $31.51$ & $+0.17$ & $34.64$ & $3.008$ & $-1.31$ & $0.40$ $(2.7)$ & $0.13$ \\
SDSS~J$091316.55+591921.6$ & $20.39$ & $-27.03$ & $3.85$ & $0.51^{+0.34}_{-0.29}$ & $0.16$ & $0.95$ & $44.34$ & $>0.71$ & $0.36$ & $31.23$ & $-0.67$ & $34.07$ & $2.717$ & $-1.76$ & $-0.09$ $(0.6)$ & $-0.32$ \\
RX~J$1028.6-0844$ & $19.14$ & $-27.95$ & $4.60$ & \nodata & $38.14$ & $179.40$ & $46.49$ & $1.40^{+0.03}_{-0.03}$ & $1.26$ & $31.65$ & $-0.30$ & $35.03$ & $3.329$ & $-1.09$ & $0.63$ $(4.3)$ &$0.34$ \\
PMN~J$1155-3107$ & $19.28$ & $-27.90$ & $6.04$ & \nodata & $5.63$ & $29.79$ & $45.34$ & $>1.33$ & $1.01$ & $31.56$ & $+0.53$ & $34.37$ & $2.727$ & $-1.36$ & $0.36$ $(2.4)$ & $0.12$ \\
SDSS~J$123503.03-000331.7$ & $20.10$ & $-27.20$ & $1.90$ & \nodata & $2.15$ & $10.57$ & $44.96$ & \nodata & $0.16$ & $30.81$ & \nodata & $33.87$ & $3.051$ & $-1.22$ & $0.39$ $(1.9)$ & $0.16$ \\
CLASS~J$1325+1123$ & $19.18$ & $-28.01$ & $1.93$ & $5.04^{+1.26}_{-1.05}$ & $1.62$ & $10.71$ & $45.29$ & $1.80^{+0.50}_{-0.40}$ & $1.03$ & $31.59$ & $-0.09$ & $34.43$ & $2.722$ & $-1.53$ & $0.19$ $(1.3)$ & $-0.05$ \\
GB~$1428+4217$ & $19.10$ & $-28.18$ & $1.39$ & \nodata & $70.07$ & $447.30$ & $46.95$ & $1.73^{+0.03}_{-0.03}$ & $1.15$ & $31.68$ & $+0.37$ & $34.87$ & $3.065$ & $-0.93$ & $0.80$ $(5.5)$ & $0.52$  \\
GB~$1508+5714$ & $19.92$ & $-27.16$ & $1.46$ & $42.06^{+0.70}_{-0.69}$ & $15.08$ & $76.68$ & $46.13$ & $1.55^{+0.05}_{-0.05}$ & $0.24$ & $30.94$ & $+0.13$ & $34.93$ & $3.871$ & $-0.96$ & $0.67$ $(3.4)$ & $0.34$ \\
GB~$1713+2148$ & $21.42$ & $-25.53$ & $5.05$ & $6.63^{+0.94}_{-0.83}$ & $2.36$ & $12.40$ & $45.29$ & $1.63^{+0.24}_{-0.23}$ & $0.13$ & $30.61$ & $-0.30$ & $35.11$ & $4.501$ & $-1.16$ & $0.42$ $(2.1)$ & $0.05$  \\
PMN~J$1951+0134$ & $19.69$ & $-27.40$ & $16.42$ & \nodata & $8.78$ & $42.79$ & $45.51$ & $1.54^{+0.33}_{-0.30}$ & $0.69$ & $31.36$ & $+0.24$ & $34.48$ & $3.036$ & $-1.23$ & $0.45$ $(3.1)$ & $0.20$ \\
\enddata
\tablenotetext{a}{The apparent $i$-band magnitude.}
\tablenotetext{b}{The absolute $i$-band magnitude, corrected for Galactic extinction.}
\tablenotetext{c}{The count rate of the \chandra-observed sources in the observed-frame soft \hbox{X-ray} band ($0.5-2.0$ keV), in units of $10^{-3}$ ${\rm s}^{-1}$.}
\tablenotetext{d}{The Galactic absorption-corrected observed-frame flux between $0.5-2.0$ keV in units of $10^{-14}$ ergs cm$^{-2}$ s$^{-1}$.}
\tablenotetext{e}{The flux density at rest-frame 2 keV, in units of $10^{-32}$ ergs cm$^{-2}$ s$^{-1}$ Hz$^{-1}$.}
\tablenotetext{f}{The logarithm of the X-ray luminosity in the rest-frame 2--10~keV band, corrected for Galactic absorption.}
\tablenotetext{g}{The X-ray power-law photon index (also see Table~\ref{cts_table}).}
\tablenotetext{h}{The flux density at rest-frame 2500~\AA\ in units of 10$^{-27}$ ergs cm$^{-2}$ s$^{-1}$ Hz$^{-1}$.}
\tablenotetext{i}{The logarithm of the monochromatic UV luminosity at rest-frame 2500~\AA.}
\tablenotetext{j}{The radio spectral index $\alpha$ between observed-frame 1.4--5 GHz, defined as $f_\nu\propto \nu^\alpha$.}
\tablenotetext{k}{The logarithm of monochromatic radio luminosity at rest-frame 5~GHz.}
\tablenotetext{l}{The logarithm of radio loudness; see \S\ref{xray} for definition.}
\tablenotetext{m}{$\Delta\alpha_{\rm ox, RQQ}$: the difference between the measured $\alpha_{\rm ox}$ and the expected $\alpha_{\rm ox}$ for RQQs with similar UV luminosity, defined by the $\alpha_{\rm ox}-L_{2500~{\rm \AA}}$ relation in equation (3) of Just et al.~(2007). The statistical significance of this difference, $\sigma$, is measured in units of the RMS $\alpha_{\rm ox}$ defined in Table 5 of Steffen et al.~(2006). The error range of \daox\ is $\sim0.003$--$0.07$.}
\tablenotetext{n}{$\Delta\alpha_{\rm ox, RLQ}$: the difference between the measured $\alpha_{\rm ox}$ and the expected $\alpha_{\rm ox}$ for RLQs with similar UV and radio luminosities, defined by the $L_{2~{\rm keV}}-L_{2500~{\rm \AA}}-L_{5~{\rm GHz}}$ relation in Table~7 of Miller et~al. (2011) for their full sample.}
\end{deluxetable}
\end{turnpage}